\pdfoutput=1
\documentclass{biophys-new}
\usepackage[utf8]{inputenc}
\usepackage[colorlinks,allcolors=cyan!70!black]{hyperref} 
\usepackage{amsmath}
\usepackage{graphics,fancyhdr,graphicx,subfigure}
\usepackage{amsfonts}
\usepackage{siunitx}
\usepackage{booktabs} %

\usepackage{lipsum}

\title{Receptor Abundance Alone Underpredicts Red Blood Cell Adhesion in Gaucher Disease}
\runningtitle{Adhesion of Gaucher Disease Red Blood Cells} %

\author[1,*]{Zhaojie Chai}
\author[2]{Pierre A. Buffet}
\author[2,*]{Melanie Franco}
\author[1,*]{George Em Karniadakis}
\runningauthor{Chai et al.} %

\affil[1]{Division of Applied Mathematics, Brown University, Providence, Rhode Island, United States}
\affil[2]{Universit\'e Paris Cit\'e, INSERM UMR\_S 1134, Biologie Int\'egr\'ee du Globule Rouge, and Laboratoire d'Excellence GR-Ex, Paris, France}

\affil[*]{Corresponding authors}

\papertype{Article}
\makeatletter
\renewcommand{\@maketitle}{%
\thispagestyle{titlepage}
{\raggedright\bfseries\sffamily\huge\@title\par}
\bigskip
{\raggedright\@author\par}%
}
\makeatother

\begin{document}
\begin{frontmatter}

\begin{abstract}
Red blood cells (RBCs) in Gaucher disease (GD) adhere abnormally to vessel walls and carry more Lutheran/basal cell adhesion molecule (Lu/BCAM), a laminin receptor. Whether receptor abundance sets this adhesion is untested. We simulated RBCs carrying stochastic Lu/BCAM--laminin bonds, calibrated a population on control donors, and predicted GD adhesion from the 1.3-fold receptor rise. On laminin and endothelium the data exceed the prediction 1.6- to 1.7-fold. Matching them requires 2.4- and 4.6-fold more binding-competent receptors, compatible with the estimated 5.7-fold rise in phosphorylated receptors. Neither a stiffer membrane nor higher bond affinity removes the shortfall at high shear stress. In sickle cell disease, given active-receptor counts, the model predicts the fall in adhesion under hydroxyurea within 1.4-fold, while total receptors rise. Adhesion likely follows receptor state more than abundance. Stiffening instead retains cells at a splenic slit.
\end{abstract}

\end{frontmatter}
\noindent\textbf{Teaser:} More Lu/BCAM receptors explain only part of Gaucher red cell adhesion. Receptor activation may supply the rest.\par\medskip

\section*{Introduction}

Gaucher disease (GD) is an autosomal recessive lysosomal storage disorder in which glucocerebrosidase deficiency leads to the accumulation of glucosylceramide in macrophages and other cell types~\cite{Stirnemann2017Pathophysiology}. It affects mainly the spleen, the liver and the skeleton~\cite{Degnan2019Imaging,hughes2019gaucher}, but red blood cells (RBCs) are affected as well and may contribute to the pathology~\cite{franco2013abnormal, dupuis2022phagocytosis}. Patients frequently present with mild anemia, and their RBCs show abnormal morphology, a membrane made more rigid by its altered lipid composition and elongated tethers under shear~\cite{franco2013abnormal,bratosin2011cytometric,chai2026multiscale,dupuis2020effects,reihani2016unexpected}. The reduced surface-to-volume ratio and the stiffer membrane hinder their passage through the interendothelial slits of the spleen~\cite{pivkin2016biomechanics,asaro2021red,chai2024silico,chai2026quantifying}.

GD RBCs also adhere abnormally. Under flow, they adhere more than control RBCs to human microvascular endothelial cells (HMEC-1) and to laminin~$\alpha$5~\cite{franco2013abnormal}. Lutheran/basal cell adhesion molecule (Lu/BCAM), the only laminin receptor of RBCs, is expressed at a higher level on GD RBCs and is more strongly phosphorylated, whereas four other adhesion receptors are unchanged~\cite{franco2013abnormal}. Binding of Lu/BCAM to laminin~$\alpha$5 is controlled by phosphorylation of its cytoplasmic tail, chiefly at the protein kinase~A site~\cite{gauthier2005protein}. Lu/BCAM also mediates abnormal RBC adhesion in sickle cell disease~\cite{hines2003novel}. Adhesion of RBCs to the vessel wall may contribute to the ischemic events of GD in the spleen and in bone~\cite{franco2013abnormal,lafforgue2006pathophysiology}.

The rise in Lu/BCAM expression on mature GD RBCs is modest, about 1.3-fold in the median~\cite{franco2013abnormal}, whereas adhesion rises up to 2.6-fold (Results), and across donors the receptor level correlates only moderately with adhesion~\cite{franco2013abnormal}. Whether the higher receptor number explains the stronger adhesion, or whether each receptor must also bind more effectively, cannot be read off the adhesion data. More receptors and bonds that resist force better both keep cells attached at higher shear stress, and they change a detachment curve in nearly the same way (Supplementary Materials).

Here we answer this question with a model calibrated on the control donors and used to predict the GD donors. We first reanalyze the donor-level data on endothelium and on laminin~$\alpha$5. We then simulate single RBCs carrying stochastic Lu/BCAM--laminin bonds in shear flow with dissipative particle dynamics (DPD)~\cite{fedosov2010multiscale,Fedosov2011Multiscale}. This gives the arrest of a cell, how strongly its bonds slow it, as a function of its receptor number and the shear stress. Averaging this response over a distribution of receptor numbers calibrated on the control data, we predict the GD curves for the measured rise in Lu/BCAM expression. The prediction falls short of the data, and the rise in receptor number that the data require lies between the measured rises in Lu/BCAM expression and in phosphorylated receptors. Neither the stiffer membrane of GD RBCs nor a higher affinity of each bond removes the shortfall at high shear stress.

\begin{figure}[!tbp]
\begin{center}
\includegraphics[width=\textwidth,height=0.965\textheight,keepaspectratio]{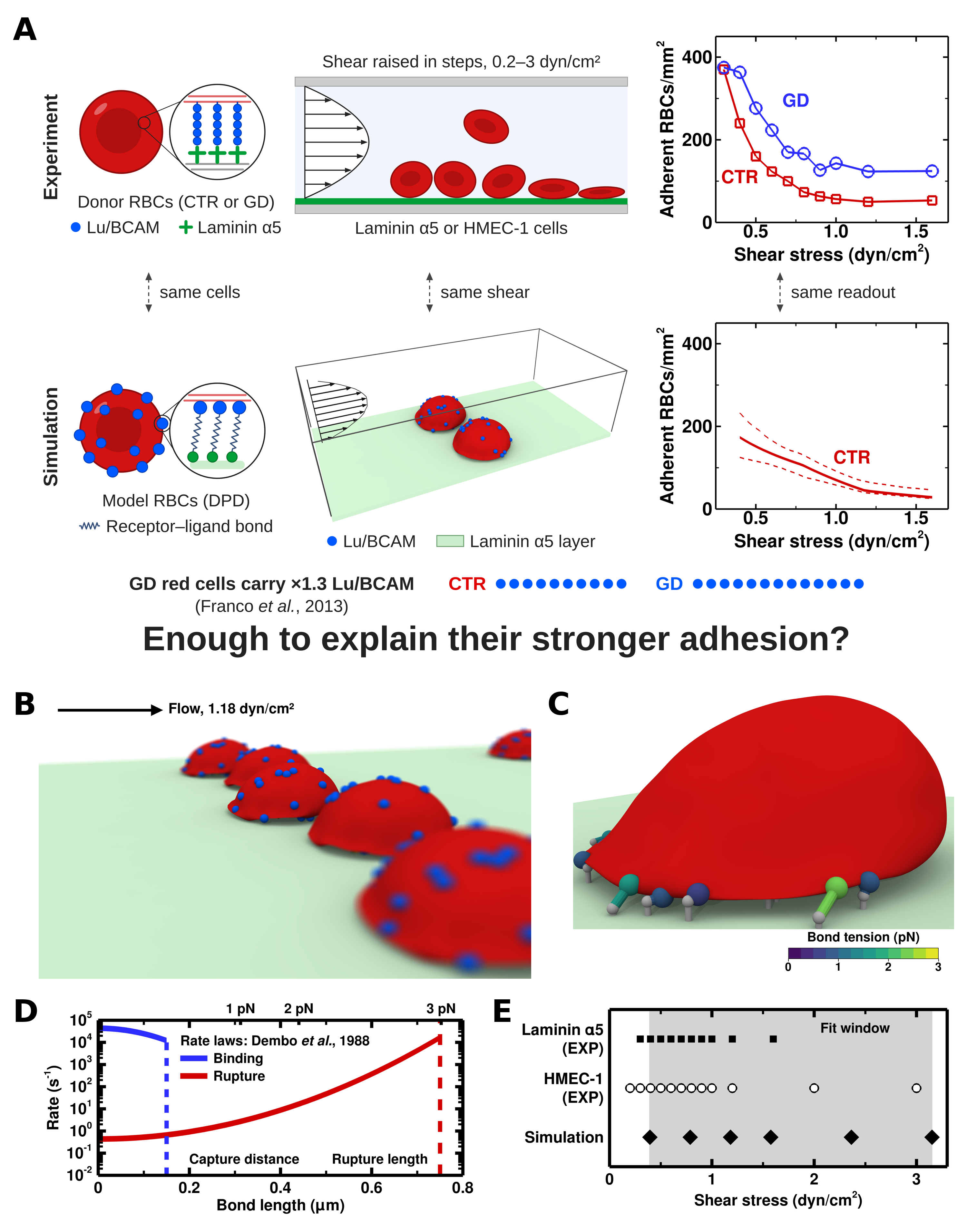}
\end{center}
\vspace{-0.15in}
\caption{\small{\bf A model of the flow-adhesion assay.}}
\label{fig:setup}
\end{figure}

\begin{figure}[!t]
\small\noindent{\bf Figure~\ref{fig:setup} (continued).}
(A) Experiment (schematic): donor RBCs (control, CTR, or Gaucher disease, GD) adhere to laminin~$\alpha$5 (green; inset: Lu/BCAM, blue) or to HMEC-1 cells in a flow chamber. The shear stress is raised in steps, and the adherent cells are counted at each step (laminin~$\alpha$5, donor medians of $n = 17$ CTR and 23 GD donors). Simulation: DPD RBCs whose Lu/BCAM receptors (blue) form spring-like bonds with sites (green) of the laminin~$\alpha$5 layer (pale green), in channel flow at comparable shear stresses. Readout: the simulated population fitted to the CTR donors (line; dashed, 68\% bootstrap band). GD RBCs carry about 1.3 times more Lu/BCAM than CTR RBCs~\cite{franco2013abnormal}. The rows of dots are a schematic of that ratio and do not show counts.
(B) The simulated chamber at 1.18~dyn/cm\textsuperscript{2}: RBCs (red) with receptors (blue) on 1/16 of their membrane vertices, over the laminin layer (pale green). The view repeats the two simulated cells periodically. The renderings draw each membrane as a smooth surface through its DPD vertices and the laminin layer as a continuous coat; its binding sites are shown only where bonded.
(C) The rear of a bound cell ($f = 1$, 2.36~dyn/cm\textsuperscript{2}): receptor--ligand bonds (tubes) and bonded receptors (spheres) colored by bond tension; gray spheres, the bound laminin sites.
(D) Kinetics of one receptor--ligand bond in the model: binding rate against the receptor--ligand distance (blue; bonds form only within 0.15~$\mu$m) and rupture rate against bond length (red; outright rupture beyond 0.75~$\mu$m)~\cite{dembo1988reaction}. The bond length is the distance between a receptor and the laminin site it is bound to. A receptor binds most readily to a site close to it, and a bond breaks faster the more it is stretched. Top axis, bond force at 3.94~pN/$\mu$m.
(E) Shear stresses of the two washout assays and of the simulations. Shaded, the shear stresses used to fit the model to the data.
\end{figure}
\section*{Results}

\subsection*{A Model of the Flow-Adhesion Assay}
We test whether the 1.3-fold higher Lu/BCAM expression of GD RBCs, measured by Franco et al.~\cite{franco2013abnormal}, accounts for their stronger adhesion with a simulated counterpart of the washout assay (Fig.~\ref{fig:setup}A). In the experiment, donor RBCs adhere to laminin~$\alpha$5 or to activated HMEC-1 cells in a flow chamber. The shear stress, the force per unit area that the flow exerts on the lower surface of the chamber where the cells adhere, is raised in steps, and the adherent cells are counted after each step. In the simulation, RBCs carrying Lu/BCAM receptors bind laminin~$\alpha$5 sites on the lower surface of a channel (Fig.~\ref{fig:setup}B, C) through stochastic bonds. Their rates of binding and rupture depend on the bond length, the distance between a receptor and its laminin site (Fig.~\ref{fig:setup}D; Materials and Methods). A cell's membrane is a mesh of 500 vertices, each carrying a receptor with probability $f$, the receptor fraction. A run lets the cells bind without flow and then applies a constant shear stress for at most 0.24~s. We call this second part the loaded phase: the flow pulls on each adherent cell, and this load is the force that its bonds must hold. For each cell we compute the arrest $A = 1 - v/v_0$, the share of the velocity $v_0$ of a non-adherent cell that the bonds take away: 0 for a free cell, 1 for a cell held in place (Fig.~\ref{fig:arrest_map}B, C). We do this at six shear stresses from 0.39 to 3.15~dyn/cm\textsuperscript{2}, which cover the shear stresses of both assays from 0.4~dyn/cm\textsuperscript{2} upward (Fig.~\ref{fig:setup}E). Averaging the arrest over a population of cells with different receptor numbers, calibrated on the control donors, gives the adherent density in cells/mm\textsuperscript{2}, the readout of the experiment.

\subsection*{GD RBCs Are Retained More Strongly during Washout}

\begin{figure}[!tbp]
\begin{center}
\includegraphics[width=\textwidth,height=0.8\textheight,keepaspectratio]{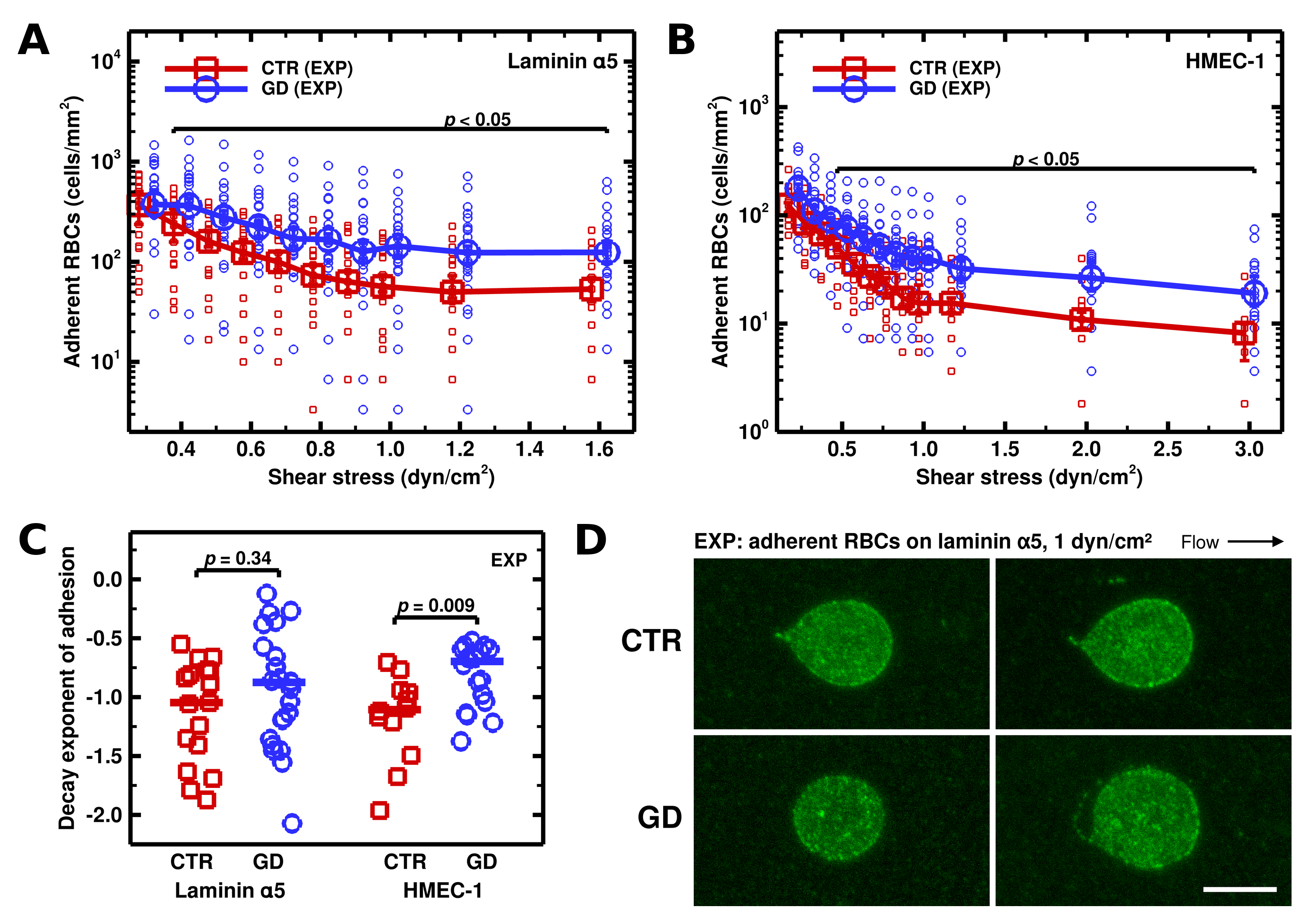}
\end{center}
\vspace{-0.15in}
\caption{\small{\bf GD RBCs adhere more than control RBCs once the shear stress of the washout is raised.}
(A, B) Adherent RBC density on laminin~$\alpha$5 (A) and on TNF$\alpha$-activated HMEC-1 endothelial cells (B) against the shear stress of the washout. EXP: small symbols, each donor (one value per donor, repeat runs averaged; the zero values, 5 in A and 4 in B, cannot be drawn on the logarithmic axis); large symbols, medians over $n = 17$ control and 23 GD donors (A) and 12 control and 18 GD donors (B; 17 GD donors at 0.2~dyn/cm\textsuperscript{2}); bars, 68\% bootstrap interval over donors. Control and GD are drawn just left and right of each shear stress. Brackets: shear stresses at which the groups differ (two-sided Mann--Whitney test at each shear stress, $p < 0.05$, not corrected for the number of shear stresses).
(C) Decay exponent of adhesion with the shear stress of the washout for each donor, the slope of log adherent density against log shear stress; bars, medians; $p$, two-sided Mann--Whitney test.
(D) Single RBCs adherent to laminin~$\alpha$5 at 1~dyn/cm\textsuperscript{2}, imaged by Lu/BCAM immunofluorescence and confocal microscopy: two cells from a control donor (top) and two from a GD patient (bottom), shown as maximum-intensity projections of the confocal stacks. These are previously unpublished images from the experiments of ref.~\cite{franco2013abnormal}, which printed one control and one GD cell of the same series; neither of those two cells is among the four shown here. Flow from left to right; all four panels are at the same magnification. The control cells carry the trailing point of the racket shape, which is more frequent in control than in GD RBCs~\cite{franco2013abnormal}. Scale bar, 5~$\mu$m.}
\label{fig:gd_experiment}
\end{figure}

Figure~\ref{fig:gd_experiment} reports the surface density of adherent RBCs on the two substrates used in the source experiments~\cite{franco2013abnormal}: TNF$\alpha$-activated HMEC-1 endothelial monolayers and microslides coated with purified laminin~$\alpha$5. The two assays bring the cells into contact with the substrate in different ways (Materials and Methods). We take this difference into account by comparing each assay only with its own controls, in the data and in the model. We analyze one value per donor at each shear stress, so each point is a donor. Five laminin runs whose preparation the experimenter questioned are excluded (Supplementary Materials). Single cells held on laminin~$\alpha$5 in these experiments, imaged by Lu/BCAM immunofluorescence, are shown in Fig.~\ref{fig:gd_experiment}D.

GD RBCs adhere more strongly than controls on both substrates, but the difference is not significant at the lowest shear stress applied. The GD/control ratios of median densities quoted here are plotted in fig.~S1A. At the lowest shear stress the ratio is 1.43 at 0.2~dyn/cm\textsuperscript{2} on endothelium ($p=0.26$) and 1.01 at 0.3~dyn/cm\textsuperscript{2} on laminin ($p=0.71$). The groups separate only as the shear stress of the washout is raised. On HMEC-1 the ratio reaches 1.46 at 0.5~dyn/cm\textsuperscript{2}, the first level at which the groups differ ($p=0.026$). It peaks at 2.56 at 1~dyn/cm\textsuperscript{2} and remains at 2.33 at 3~dyn/cm\textsuperscript{2}. On laminin it reaches 1.51 at 0.4~dyn/cm\textsuperscript{2} ($p=0.035$) and 2.53 at 1~dyn/cm\textsuperscript{2}. Every level from these onsets upward is significant when each shear stress is tested on its own (Fig.~\ref{fig:gd_experiment}A, B). These tests are not corrected for the number of shear stresses, but a single test per substrate gives the same answer. For it we average the data of Fig.~\ref{fig:gd_experiment}A and B over the shear stresses above the lowest, one mean per donor. GD donors then retain more cells than controls on laminin ($p=0.016$) and on HMEC-1 ($p=0.026$). GD cells are therefore not captured more often when they first reach the substrate. They are held more strongly once the shear stress rises.

On endothelium GD cells also resist the washout better: their adhesion falls more slowly as the shear stress rises. A per-donor regression of $\log$ density on $\log$ shear stress gives a decay exponent of $-1.11$ for controls and $-0.70$ for GD ($p=0.009$). On purified laminin the two exponents do not differ significantly, $-1.05$ and $-0.88$ ($p=0.34$; Fig.~\ref{fig:gd_experiment}C). These two summaries are different statistics and need not agree: the ratio of medians compares groups at a fixed shear stress, whereas the exponent is fitted within each donor across shear stresses. A shallower decay alone does not tell whether GD cells carry more receptors or bonds that resist force better. In a reduced Bell description of the detachment curve, both changes move its height and its slope in almost the same direction (Supplementary Materials). The decay exponent is therefore no evidence about bond mechanics without an independent measurement of receptor density.

The effect is ligand-specific. In the source experiments neither GD nor control RBCs adhered appreciably to fibronectin, whereas GD RBCs adhered more than controls to laminin~$\alpha$5 in the same paired comparison~\cite{franco2013abnormal}. Laminin~$\alpha$5 is expressed at the surface of TNF$\alpha$-activated HMEC-1 and is known to promote RBC adhesion to these cells~\cite{franco2013abnormal}, so the endothelial and purified-protein assays plausibly probe the same pathway. CD36, CD49d, CD29 and CD242 do not differ in surface level between the groups~\cite{franco2013abnormal}, which leaves Lu/BCAM as the receptor whose level differs. A change of state in the others is not excluded. We represent adhesion below with a single species of stochastic bond.

One limitation of the source dataset bears on the laminin measurements. The purified laminin was not drawn from a single lot, and the lot is recorded for too few donors to test for a lot effect (Supplementary Materials). Lot-to-lot variability cannot be excluded as a contributor to the spread between donors on laminin. It cannot, however, account for the endothelial measurements, which use no purified laminin coating and which carry the same effect at a comparable magnitude. The two substrates support each other on this point.

\subsection*{Receptor Number Decides between Arrest and Sweep}
\begin{figure}[!tbp]
\begin{center}
\includegraphics[width=\textwidth,height=0.8\textheight,keepaspectratio]{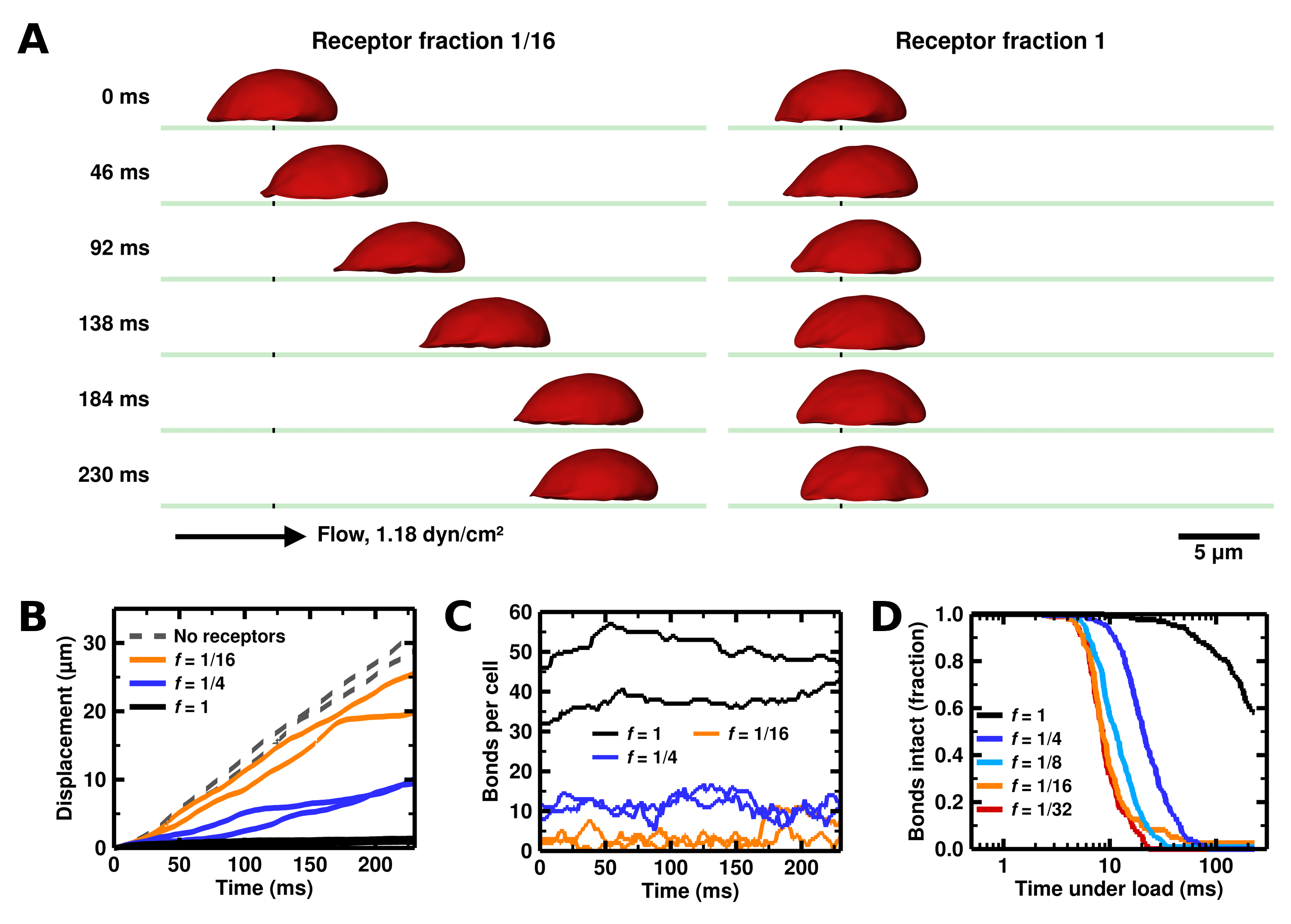}
\end{center}
\vspace{-0.15in}
\caption{\small{\bf Receptor number decides between arrest and sweep.}
(A) Side views of the same simulated cell with receptors on 1/16 ($f = 1/16$, left) and on all ($f = 1$, right) of its membrane vertices at 1.18~dyn/cm\textsuperscript{2}, every 46~ms after the flow is switched on. Tick, starting position; pale green, the laminin layer.
(B) Displacement against time at 1.18~dyn/cm\textsuperscript{2} for cells that cannot bind (dashed) and with $f = 1/16$, 1/4 and 1; one curve per cell, two cells per run.
(C) Number of receptor--ligand bonds of each cell against time, for $f = 1/16$, 1/4 and 1 at 1.18~dyn/cm\textsuperscript{2}.
(D) Fraction of bonds still intact against the time they have been under load, at 1.18~dyn/cm\textsuperscript{2} (Kaplan--Meier estimate; both cells of a run pooled; $n = 106$--253 bonds per $f$).}
\label{fig:single_cell}
\end{figure}

At a given shear stress, the number of receptors decides whether a cell is swept along the lower surface of the channel or held (Fig.~\ref{fig:single_cell}). At 1.18~dyn/cm\textsuperscript{2}, a cell with receptors on 1/16 of its membrane vertices ($f = 1/16$) slides 19.8~$\mu$m during the loaded phase, whereas the same cell with $f = 1$ moves 1.4~$\mu$m and is held (Fig.~\ref{fig:single_cell}A). In that time, cells that cannot bind travel 29--31~$\mu$m, cells with $f = 1/16$, 20--26~$\mu$m, and cells with $f = 1/4$, about 10~$\mu$m (Fig.~\ref{fig:single_cell}B). A sliding cell holds few bonds at a time, a median of 2--3 at $f = 1/16$ and 11 at $f = 1/4$, against 38--51 at $f = 1$ (Fig.~\ref{fig:single_cell}C). Its bonds are also short-lived. In the simulations, half of them break within 8--21~ms under load for $f$ between 1/32 and 1/4, whereas at $f = 1$ more than half survive the whole loaded phase (Fig.~\ref{fig:single_cell}D).

\subsection*{Receptor Number Matters More at Higher Shear Stress}
\begin{figure}[!tbp]
\begin{center}
\includegraphics[width=\textwidth,height=0.8\textheight,keepaspectratio]{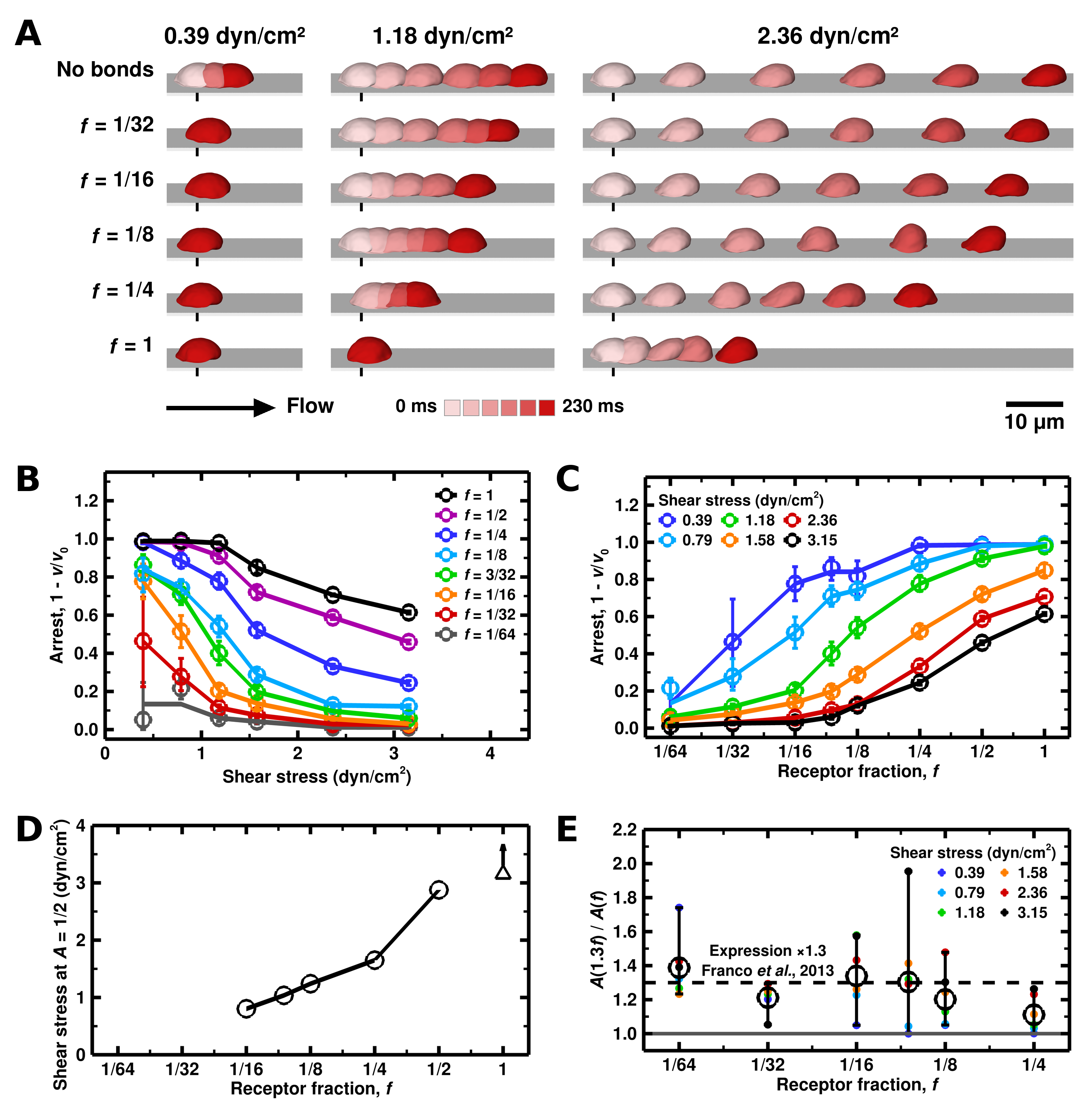}
\end{center}
\vspace{-0.15in}
\caption{\small{\bf Receptor number matters more at higher shear stress.}
(A) Simulated cells at 0.39, 1.18 and 2.36~dyn/cm\textsuperscript{2} (columns), without bonds (top row; the non-adherent control) and at five of the receptor fractions of B, $f = 1/32$ to 1 (rows), seen from the front and slightly above. Each condition shows one simulated cell, drawn every 46~ms over the first 230~ms of flow, pale to red. A copy within 2.5~$\mu$m of the next one drawn is left out, so a cell that arrests shows as one red cell; the sideways drift of the cell (up to 2.9~$\mu$m) is removed. Black ticks, its starting position.
(B) Arrest $A = 1 - v/v_0$ of single cells against shear stress for eight receptor fractions $f$; $v_0$ is the velocity of non-adherent control cells. Symbols: mean of $n = 12$ cells (6 runs $\times$ 2 cells; $n = 24$ at $f = 1/64$); bars, 68\% bootstrap interval. Lines: the same table made non-increasing in shear stress and non-decreasing in $f$ (Materials and Methods), as used by the population model.
(C) The same arrest against $f$ at the six simulated shear stresses (symbols and bars as in B; lines, the monotone table).
(D) Shear stress at which a cell loses half of its free speed ($A = 1/2$) against $f$. Open triangle, lower bound ($A > 1/2$ up to 3.15~dyn/cm\textsuperscript{2}).
(E) Gain in single-cell arrest from a 1.3-fold rise in receptors, $A(1.3f)/A(f)$, against $f$. Dots, the six shear stresses; circles, their geometric mean; bars, their range. Dashed line, 1.3.}
\label{fig:arrest_map}
\end{figure}

Over the full grid of eight receptor fractions and six shear stresses, the arrest $A$ (0 for a free cell, 1 for a cell held in place) falls as the shear stress rises, and the range of shear stresses over which a cell is held widens with its receptor fraction $f$ (Fig.~\ref{fig:arrest_map}A--C). At 1.18~dyn/cm\textsuperscript{2}, a cell with $f = 1/16$ is carried along at 80\% of the velocity of a non-adherent cell ($A = 0.20$), whereas a cell with $f = 1$ stays in place ($A = 0.98$). The shear stress at which a cell loses half of its free speed grows about as $f^{0.59}$, from 0.80~dyn/cm\textsuperscript{2} at $f = 1/16$ to 2.88~dyn/cm\textsuperscript{2} at $f = 1/2$ (Fig.~\ref{fig:arrest_map}D). How much a cell gains from more receptors depends on the shear stress. At 0.39~dyn/cm\textsuperscript{2}, $A$ is already 0.46 at $f = 1/32$ and 0.98 at $f = 1/4$, close to its ceiling. At 3.15~dyn/cm\textsuperscript{2} it stays below 0.15 up to $f = 1/8$ and reaches 0.61 at $f = 1$. From $f = 1/16$ upward, a given relative rise in receptor number raises the arrest proportionally more at high shear stress. For a 1.3-fold rise, the gain $A(1.3f)/A(f)$, new arrest over old, has a geometric mean of 1.26 over the six fractions up to $f = 1/4$ and the six shear stresses. At $f = 1/16$ the gain is 1.05 at 0.39~dyn/cm\textsuperscript{2} and 1.23--1.58 above (Fig.~\ref{fig:arrest_map}E). The measured GD/control ratio also grows with the shear stress (fig.~S1A). Whether a 1.3-fold rise in receptor number makes it grow enough is the question of the population analysis below.

\subsection*{Where the Force Sits and Where Bonds Fail}
\begin{figure}[!tbp]
\begin{center}
\includegraphics[width=\textwidth,height=0.8\textheight,keepaspectratio]{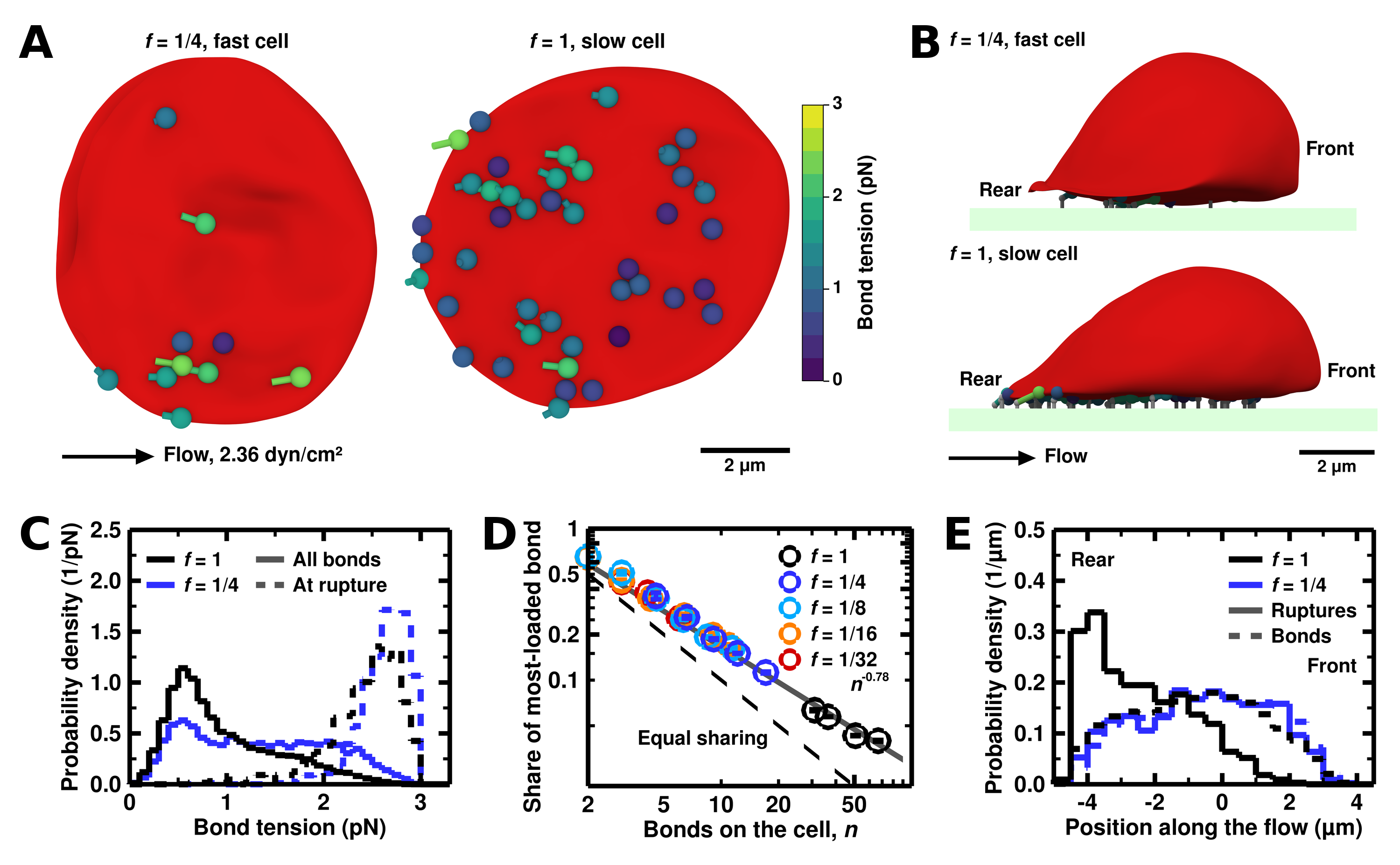}
\end{center}
\vspace{-0.15in}
\caption{\small{\bf Where the force sits and where bonds fail.}
(A) Contact zone seen from below at 2.36~dyn/cm\textsuperscript{2} for a fast cell ($f = 1/4$) and a slow cell ($f = 1$), each at the frame with its median number of bonds. Bonds and bonded receptors are colored by bond tension; unbonded receptors and ligand sites are not drawn; flow to the right. Scale bar, 2~$\mu$m.
(B) The same cells and frames seen from the side (flow to the right, rear on the left); pale green, the laminin layer; gray, the bound laminin sites.
(C) Distribution of bond tension at 2.36~dyn/cm\textsuperscript{2} over all bonds in all frames (solid) and at rupture (dashed); $f = 1/4$: $n = 4461$ bond-frames and 910 ruptures; $f = 1$: 19,194 and 657.
(D) Share of the total bond tension carried by the most-loaded bond against the number of bonds $n$ on the cell (means of frames binned in $n$ $\pm$ SEM; bins with at least 10 frames). Dashed line, equal sharing ($1/n$); gray line, the fit $n^{-0.78}$ over all $f$ and shear stresses (5489 frames).
(E) Position along the flow, relative to the cell center, of the bonds (dashed) and of the bonds that rupture (solid), at 2.36~dyn/cm\textsuperscript{2}.}
\label{fig:bonds}
\end{figure}

To see how bonds hold a cell, we reran 15 conditions with every bond recorded (Materials and Methods) and compared a fast cell ($f = 1/4$) with a slow one ($f = 1$) at 2.36~dyn/cm\textsuperscript{2} (Fig.~\ref{fig:bonds}A, B). Most bonds sit in the rear half of the contact zone. The median bond carries 1.3~pN in the fast cell and 0.8~pN in the slow one, and bonds break at 2.5--2.7~pN, two to three times as much (Fig.~\ref{fig:bonds}C). The force is not shared equally. The share carried by the most-loaded bond, the bond under the largest force, falls as $n^{-0.78}$ with the number of bonds $n$, between equal sharing ($n^{-1}$) and peeling, in which the share does not fall with $n$ (Fig.~\ref{fig:bonds}D). The two cells lose their bonds in different places (Fig.~\ref{fig:bonds}E). The slow cell holds 63\% of its bonds in the rear half but has 93\% of its ruptures there, 31\% within the rearmost micrometer. The fast cell loses its bonds about where it holds them (57\% of the bonds and 60\% of the ruptures in the rear half).

\subsection*{From Single Cells to Donor Populations}
\begin{figure}[!tbp]
\begin{center}
\includegraphics[width=\textwidth,height=0.8\textheight,keepaspectratio]{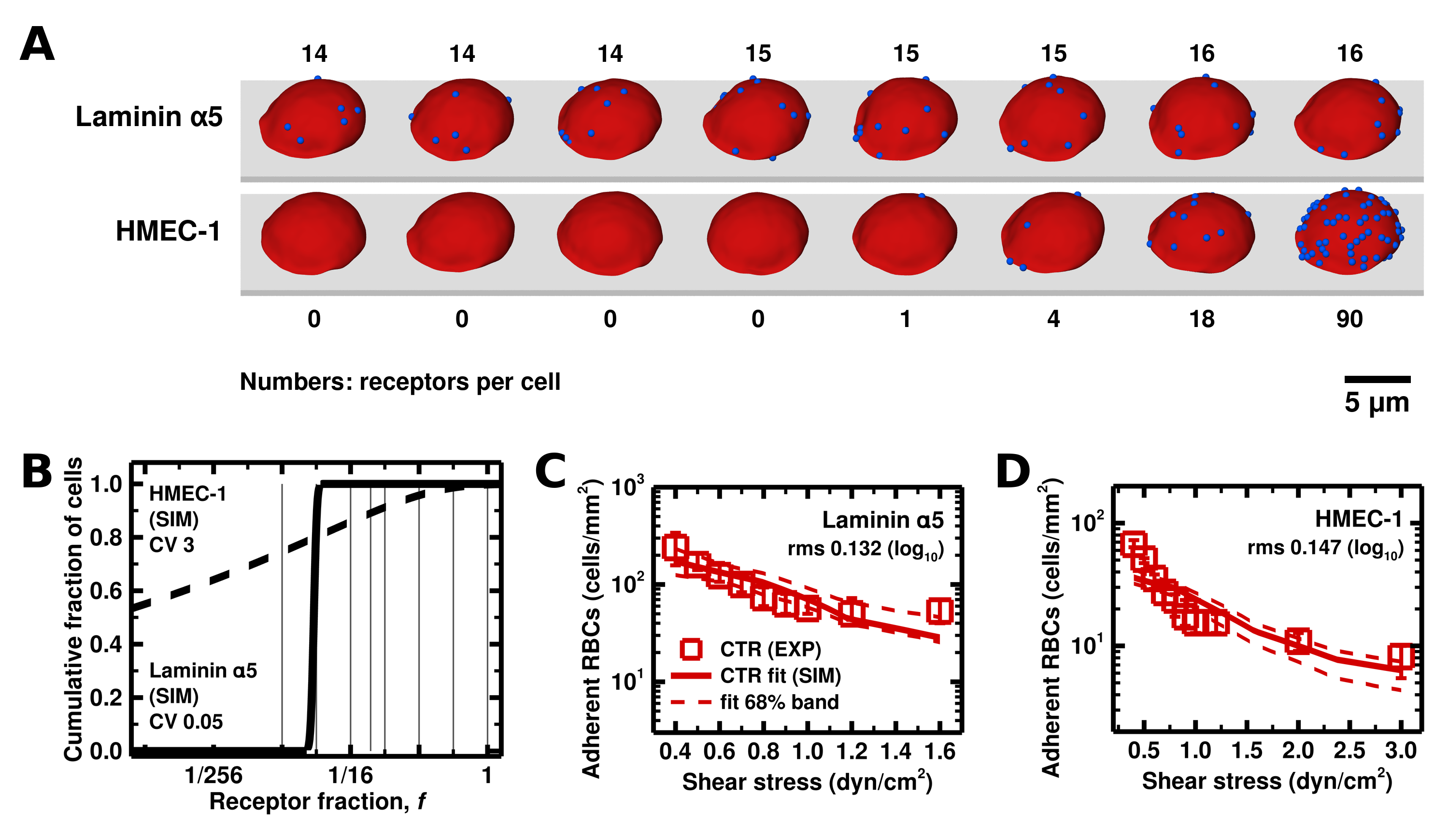}
\end{center}
\vspace{-0.15in}
\caption{\small{\bf From single cells to donor populations: calibration on the control donors.}
(A) The fitted control populations drawn on the model cell: eight cells per substrate carrying $500f$ receptors (blue), $f$ at the octile midpoints of the distributions in B, on laminin~$\alpha$5 (top) and HMEC-1 (bottom); numbers, receptors per cell. The widths of the two distributions are the bounds of the range searched: the curves do not determine them (fig.~S2A). The shapes are snapshots of the model cell at 1~dyn/cm\textsuperscript{2}, the receptors placed on random vertices.
(B) Fitted distribution of the receptor fraction $f$ among control cells, as the cumulative fraction of cells: a gamma distribution with the mean and CV fitted to the control donors of each assay (solid, laminin~$\alpha$5; dashed, HMEC-1). Gray lines mark the eight simulated fractions.
(C, D) Adherent control RBCs against shear stress on laminin~$\alpha$5 (C; 17 donors) and HMEC-1 (D; 12 donors). Symbols, donor medians with their 68\% bootstrap interval; lines, the population model $N = c\,\mathbb{E}_f[A(\tau_w; f)]$ with its 68\% bootstrap band (500 resamples of donors and simulated cells).}
\label{fig:calibration}
\end{figure}

Because the receptor number may vary between cells, we describe the RBCs of each donor group by a gamma distribution of $f$. We compute the adherent density as the population average of $A$, $N = c\,\mathbb{E}_f[A(\tau_w; f)]$, where the scale $c$ is the density of cells in contact with the substrate (Eq.~\ref{eq:population}; Materials and Methods). Fitted to the controls, the model follows the median control curve with an rms deviation in $\log_{10}$ density of 0.132 on laminin and 0.147 on HMEC-1 (Fig.~\ref{fig:calibration}C, D). The mean receptor fraction is 0.030 on laminin (68\% interval 0.026--0.066) and 0.038 on HMEC-1 (0.031--0.070), close to 1/32, the second smallest simulated fraction (Fig.~\ref{fig:calibration}A, B). The width of the distribution is not determined on either substrate. On laminin its coefficient of variation (CV) settles at the lower bound of the search range, 0.05, a population of nearly identical cells, and fixing it at 0.2, 0.5 or 1.0 changes the rms deviation by at most 0.006. On HMEC-1 it settles at the upper bound, 3, a broad distribution in which 75\% of the cells lie below $f = 1/64$, where $A$ is interpolated linearly to zero (Materials and Methods). These cells supply 5\% of the predicted adherent density at 0.39~dyn/cm\textsuperscript{2} and 2--6\% at the higher simulated shear stresses. Fixing the CV at 0.2, 0.5 or 1.0 raises the rms deviation on HMEC-1 only to 0.151--0.168 (fig.~S2).

\subsection*{The Measured Rise in Lu/BCAM Underpredicts GD Adhesion}

In the model, the measured rise in Lu/BCAM predicts only part of the adhesion of GD RBCs. To match the data, a GD cell would need 2.4 (laminin) to 4.6 (HMEC-1) times the binding-competent receptors of a control cell. This is compatible with the estimated 5.7-fold rise in phosphorylated receptors.

\begin{figure}[!tbp]
\begin{center}
\includegraphics[width=\textwidth,height=0.8\textheight,keepaspectratio]{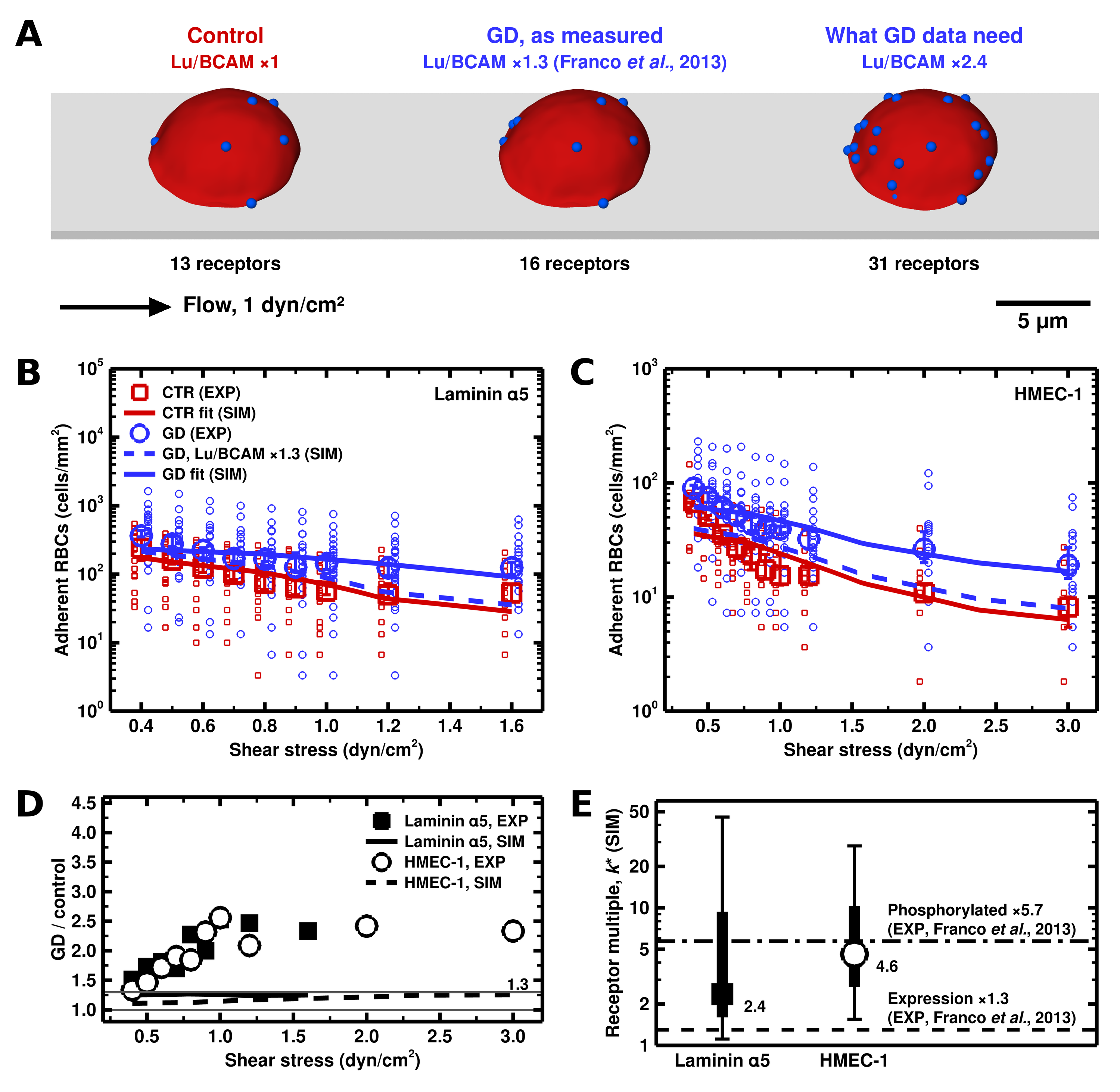}
\end{center}
\vspace{-0.15in}
\caption{\small{\bf A 1.3-fold increase in Lu/BCAM falls short of the adhesion of GD RBCs in the model.}
(A) The model cell at 1~dyn/cm\textsuperscript{2}, 115~ms after the flow starts, with the mean receptor fraction of the control fit on laminin~$\alpha$5 ($f = 0.030$; 13 receptors). Also shown: the fraction raised 1.3-fold, as measured for Lu/BCAM on GD RBCs (16 receptors), and 2.4-fold (31 receptors), the multiple $k^*$ that brings the model to the GD data on laminin~$\alpha$5 (E). Each of the 500 membrane vertices carries a receptor with probability $f$, so these counts are single draws around means of 15, 19 and 35.
(B, C) Adherent RBC density on laminin~$\alpha$5 (B) and HMEC-1 (C). EXP: small symbols, each donor, as in Fig.~\ref{fig:gd_experiment}A-B; large symbols, medians of $n = 17$ control and 23 GD donors (B) and of 12 and 18 (C); bars, 68\% bootstrap interval over donors. SIM: population with gamma-distributed $f$ fitted to the controls (red line); GD predicted by raising the mean $f$ 1.3-fold with nothing refitted (blue dashed line); GD fitted with the mean and CV of $f$ free (blue solid line).
(D) GD/control ratio of adherent density against shear stress. Symbols, ratio of the GD and control donor medians; lines, the ratio predicted for a 1.3-fold rise in Lu/BCAM. Gray lines, 1.3 and no change.
(E) Receptor multiple $k^*$ that brings the model to the GD data, with 68\% (thick) and 95\% (thin) bootstrap intervals over donors and simulation runs. Dashed, the measured 1.3-fold rise in Lu/BCAM expression; dash-dot, the 5.7-fold rise in phosphorylated Lu/BCAM per cell, $1.3 \times 4.4$~\cite{franco2013abnormal}.}
\label{fig:shortfall}
\end{figure}

Franco et al. found 1.3 times more Lu/BCAM on mature GD RBCs than on control RBCs~\cite{franco2013abnormal}. Multiplying the receptor fraction of every control cell by 1.3, with the cell density $c$ of the controls, predicts the GD curve without any fit to GD data (Fig.~\ref{fig:shortfall}B, C, dashed blue lines). The predicted rise in adherent density stays below 1.3-fold, because at these receptor fractions $A$ grows on average less than in proportion to $f$. On HMEC-1 it grows with the shear stress, as the single-cell response implies, from 1.11-fold at 0.4~dyn/cm\textsuperscript{2} to 1.25-fold at 3~dyn/cm\textsuperscript{2}. On laminin, where nearly all control cells lie close to $f = 1/32$, it stays between 1.24- and 1.27-fold from 0.4 to 1.6~dyn/cm\textsuperscript{2}. The measured GD/control ratio grows much more, from 1.34 at 0.4~dyn/cm\textsuperscript{2} to 2.1--2.6 from 0.9~dyn/cm\textsuperscript{2} upward on HMEC-1, and from 1.51 at 0.4~dyn/cm\textsuperscript{2} to 2.0--2.5 from 0.8~dyn/cm\textsuperscript{2} upward on laminin (Fig.~\ref{fig:shortfall}D). Averaged geometrically over the fitted shear stresses, the GD medians therefore exceed the prediction 1.70-fold on HMEC-1 and 1.60-fold on laminin. We call this ratio the shortfall.

The 68\% bootstrap intervals of these ratios (10,000 replicates of donors and whole simulation runs; Materials and Methods) exclude agreement on both substrates (1.41--2.15 and 1.18--2.15). The 95\% intervals are 1.14--2.69 on HMEC-1 and 0.88--3.12 on laminin (fig.~S1B). A percentile bootstrap is unreliable in the tail, so we test the shortfall by relabelling whole donor curves between the groups, which is exact under exchangeability and preserves the correlation between shear stresses (Materials and Methods). The one-sided permutation $P$ is 0.014 on HMEC-1 and 0.046 on laminin, two-sided 0.032 and 0.085: the shortfall has the same size on both substrates, and we take it as established on HMEC-1 and as marginal on laminin. The uncertainty of the shortfall comes mainly from the spread between donors and less from the simulations (table~S1).

Two assumptions of the prediction matter most, and the Supplementary Materials test both. First, no data calibrate the model's response to receptor number, and the one measurement of it is steeper. Over the fitted shear stresses a 1.3-fold rise gives 1.15 times the adhesion on HMEC-1 and 1.26 times on laminin in the model. The one dataset in which adhesion and Lu/BCAM level are given patient by patient, in polycythemia vera, gives 1.41 times (95\% interval 1.32--1.56)~\cite{brusson2018impact}. With that dependence in place of the model's, the shortfall is 1.39-fold on HMEC-1 and 1.43-fold on laminin, significant on neither. Second, GD cells are given the control density $c$ of cells in contact, which multiplies the whole predicted curve. The significance on HMEC-1 rests on this choice. It holds for a GD density up to 1.21 times that of the controls, whereas at the lowest shear stress, below the fitted range, the GD medians exceed the control medians 1.43-fold. The 1.3 itself is a ratio of donor medians, with a 95\% interval of 1.12--1.52 when the donors of the source figure are resampled.

The receptor rise that the GD data require, $k^*$, is the single factor by which the receptor fraction of every control cell must be multiplied for the model to fit the GD curve. It is 2.4 on laminin (68\% interval 1.6--9.4) and 4.6 on HMEC-1 (2.7--10; Fig.~\ref{fig:shortfall}A, E). The data bound it only from below, and on laminin its 95\% interval includes 1.3 (Supplementary Materials). The ratio of phosphorylated to total Lu/BCAM is 4.4-fold higher in GD RBCs~\cite{franco2013abnormal}, so a GD cell carries $1.3 \times 4.4 = 5.7$ times the phosphorylated receptors of a control cell. This lies inside both 68\% intervals, although the 5.7 is itself uncertain (Supplementary Materials). A different width of the receptor distribution does not replace the rise in its mean. Letting the width of the GD distribution vary as well as the mean reproduces both GD curves (Fig.~\ref{fig:shortfall}B, C, solid blue lines), but keeping the 1.3-fold mean and changing only the width does not (Supplementary Materials). Within the model, the measured rise in Lu/BCAM expression thus probably accounts for only part of the stronger adhesion of GD RBCs.

\subsection*{The Gap between Prediction and Experiment Persists}
So far both groups carry the membrane of healthy RBCs. GD RBCs are less deformable~\cite{franco2013abnormal}, and our earlier model of moderate GD gives their membrane a shear modulus of 42.6~$\mu$N/m, nine times the healthy 4.73~$\mu$N/m~\cite{chai2024silico}. With this shear modulus, the same cell, with the same receptors, thermal bath and flow, holds fewer bonds (24 of 24 cells; median ratio 0.77; Fig.~\ref{fig:alternatives}A). Its bonds also break sooner (median lifetime under load 0.79 times as long at 1.18 and 2.36~dyn/cm\textsuperscript{2}; Fig.~\ref{fig:alternatives}A). Over the 48 conditions of the grid, the stiffer membrane lowers the arrest slightly (mean change $-0.022$; lower at 28 conditions and higher at 14), and raises it on average only at the lowest shear stress (Fig.~\ref{fig:alternatives}B). The 1.3-fold prediction made with the GD membrane is therefore lower than with the healthy one. The GD data exceed this prediction 2.24-fold on laminin and 1.88-fold on HMEC-1, against 1.67- and 1.68-fold with the healthy membrane always on the same two spots of the substrate (fig.~S3C; Materials and Methods). With the GD membrane, GD cells would need 3.3 and 4.8 times the receptors of a control cell, against 3.2 and 3.9 with the healthy one (Fig.~\ref{fig:alternatives}C). A model of adherent sickle cells likewise found earlier detachment at a higher shear modulus and the same bond stiffness~\cite{Deng2019Quantifying}.

The shortfall of a uniform 1.3-fold increase persists whatever width is assumed for both groups and under the other choices of the analysis (Supplementary Materials, fig.~S4). With the CV of both groups fixed at 0.2, 0.5 or 1.0, the GD data exceed the prediction 1.57- to 1.61-fold on laminin and 1.72- to 1.74-fold on HMEC-1, and 5.8--8.0\% and 0.6--1.0\% of the replicates reach the data. A lognormal distribution gives 1.60- and 1.70-fold. Leaving out the eight non-adherent control cells that crept at the lowest shear stress (Materials and Methods) gives 1.74- and 1.73-fold, with 3.8\% and 0.4\% of the replicates reaching the data. With the arrest table as simulated, the shortfall is 1.62- and 1.72-fold, with 6.8\% and 0.6\% of the replicates reaching the data.

Two further changes to the model leave the shortfall in place (fig.~S3C). With a bond twice as stiff ($k_s = 100$, 7.9~pN/$\mu$m), the GD data exceed the 1.3-fold prediction 1.81-fold on laminin and 1.70-fold on HMEC-1. With one third of the body force on the membrane vertices, they exceed it 1.64- and 1.65-fold. In all four models run on those two spots, the 68\% interval of this ratio lies above 1, and on HMEC-1 so does the 95\% interval (lowest bound 1.06). The receptor rise that GD would need stays between 2.6 and 5.6. The lower deformability of GD RBCs does not contribute to their stronger adhesion in the model. If anything it works against it. None of the other choices tested closes the gap between prediction and experiment.

\subsection*{Bond Affinity Acts at Low Shear Stress and Receptor Number at High}
\begin{figure}[!tbp]
\begin{center}
\includegraphics[width=\textwidth,height=0.88\textheight,keepaspectratio]{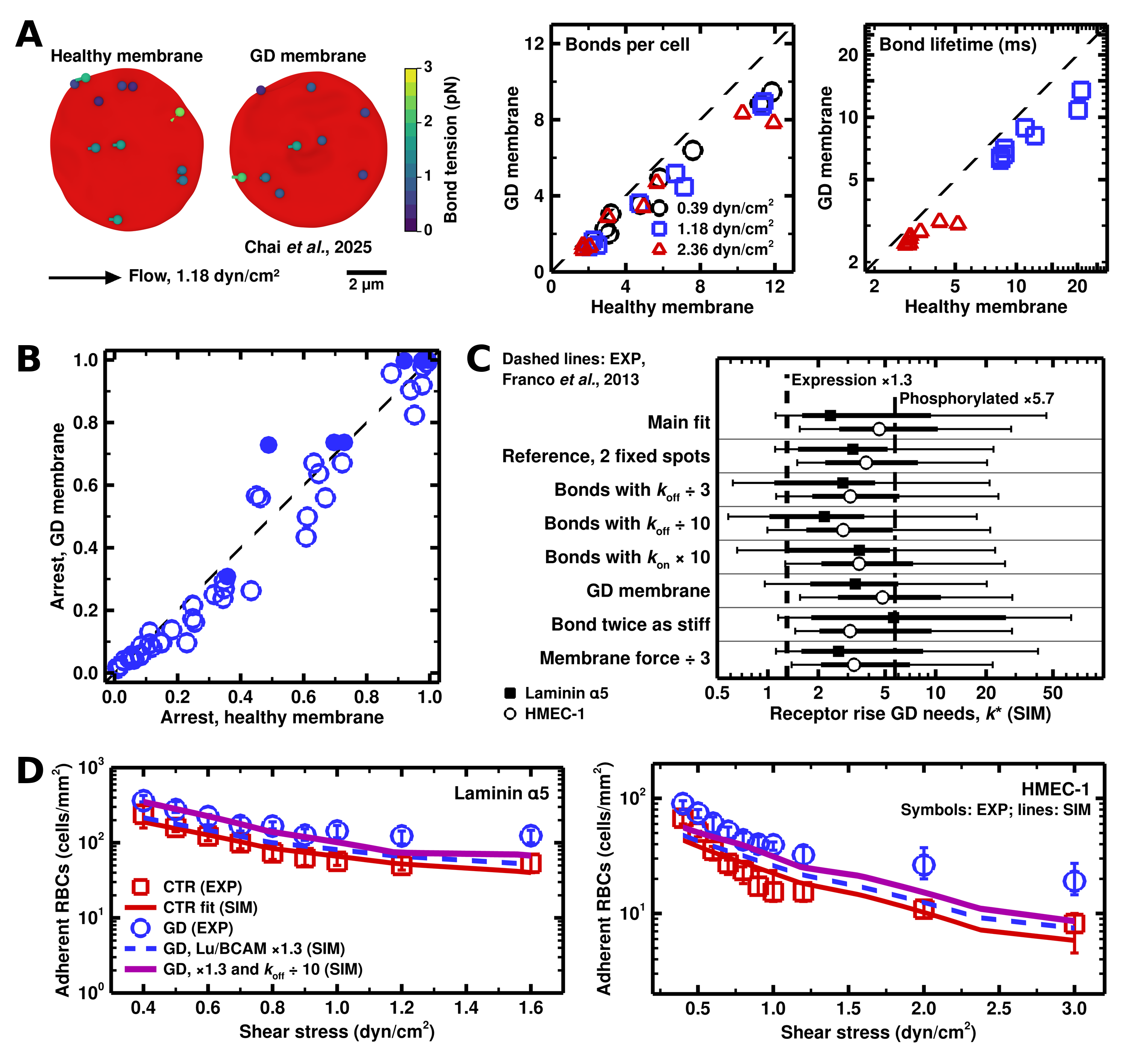}
\end{center}
\vspace{-0.15in}
\caption{\small{\bf Neither the GD membrane nor a higher bond affinity closes the gap between prediction and experiment.}
(A) Bonds with the membrane of GD RBCs. Left: contact zone of the same cell with the healthy membrane (shear modulus 4.73~$\mu$N/m) and with the GD membrane (42.6~$\mu$N/m), at $f = 1/4$ and 1.18~dyn/cm\textsuperscript{2}, drawn as in Fig.~\ref{fig:bonds}A; scale bar, 2~$\mu$m. Middle: bonds per cell averaged over the loaded phase, GD against healthy membrane (24 cells: $f = 1/32$ to 1/4 at three shear stresses, two cells per run). Right: Kaplan--Meier median bond lifetime at 1.18 and 2.36~dyn/cm\textsuperscript{2} (16 cells). Dashed lines, equal values.
(B) Arrest at each of the 48 simulated conditions (8 receptor fractions $\times$ 6 shear stresses) with the GD membrane against the healthy membrane; 4--8 cells per condition and membrane. Dashed line, no change. Filled symbols, 0.39~dyn/cm\textsuperscript{2}.
}
\label{fig:alternatives}
\end{figure}

A change in receptor state could instead raise the bond affinity. We lowered the unstressed off-rate $k_{\text{off}}^0$ threefold and tenfold, and raised the on-rate prefactor $k_{\text{on}}^0$ tenfold, on the full grid (Fig.~\ref{fig:alternatives}; Materials and Methods). A tenfold lower off-rate raises the arrest by 0.09 on average at 0.39 and 0.79~dyn/cm\textsuperscript{2}, but by only 0.03 and 0.02 at 2.36 and 3.15~dyn/cm\textsuperscript{2} (fig.~S3A). A threefold lower off-rate has about two thirds of this effect. It adds few bonds, 7.5\% over all runs, because binding is nearly saturated: a cell has about $50f$ live bonds (fig.~S3B). At 0.39~dyn/cm\textsuperscript{2}, most bonds dissociate thermally at low force, and with the lower off-rate a cell loses 0.9 instead of 3.6 bonds during the loaded phase. At 2.36 and 3.15~dyn/cm\textsuperscript{2}, the bonds are pulled close to their rupture length. With the lower off-rate, 65--73\% of the ruptures occur there, and the median rupture force rises only from 2.7 to 3.0~pN. The gain from a higher affinity therefore fades with shear stress, whereas the excess of GD adhesion grows with it (fig.~S1A; on laminin only in the central values, table~S2).

With the 1.3-fold rise, the tenfold lower off-rate brings the model to the GD data on laminin within the 68\% interval (GD/model 1.20, 0.79--1.84) but not on HMEC-1 (1.41, 1.08--1.80; Fig.~\ref{fig:alternatives}D). On both substrates it falls short at high shear stress: on laminin the ratio is 0.95--1.20 at 0.4--0.9 but 1.41--1.84 at 1.0--1.6~dyn/cm\textsuperscript{2}, and on HMEC-1 1.7 and 2.2 at 2 and 3~dyn/cm\textsuperscript{2}. A bond less sensitive to force ($\sigma_{\text{off}}/4$) gives ratios of 0.97 (68\% interval 0.68--1.56) on laminin and 1.26 (0.97--1.62) on HMEC-1, but falls short at the higher shear stresses. With a rupture length one third longer as well, the ratios are 0.73 (0.55--1.26) and 1.07 (0.81--1.36), and on HMEC-1 1.12 and 1.35 at 2 and 3~dyn/cm\textsuperscript{2} (Supplementary Materials). A tenfold faster on-rate raises the arrest by 0.08 on average at 0.39~dyn/cm\textsuperscript{2} and by at most 0.01 above. The fitted $k^*$ is 2.2--3.5 with the rate changes of Fig.~\ref{fig:alternatives}C and 1.9 and 2.3 with the less force-sensitive bond, against 3.2 and 3.9 with default bonds on the same spots (lower 68\% bounds on laminin: 1.0--1.3, 0.8 and 1.5). With each changed bond in both groups and the controls refitted, the default shortfall returns (1.57--1.81 and 1.49--1.74, against 1.67 and 1.68; table~S3). A higher affinity thus acts at low shear stress. At fixed rupture length, the excess of GD adhesion at high shear stress requires more receptors.

\begin{figure}[!t]
\small\noindent{\bf Figure~\ref{fig:alternatives} (continued).}
(C) $k^*$ on laminin~$\alpha$5 and HMEC-1 for the model fitted to the runs of Fig.~\ref{fig:arrest_map} (main fit), for reference runs in which the two cells of every run settle over the same two spots of the substrate, and for models run on those spots (Materials and Methods): bonds with $k_{\text{off}}^0/3$, with $k_{\text{off}}^0/10$ or with $10\,k_{\text{on}}^0$, the GD membrane, a bond twice as stiff (7.9 instead of 3.9~pN/$\mu$m), and one third of the body force on the membrane. Dashed and dash-dot lines as in Fig.~\ref{fig:shortfall}E. Thick and thin bars are 68\% and 95\% bootstrap intervals: 10,000 resamples of donors and runs for the main fit, 500 of donors and simulated cells for the three rate changes, 5000 for the other rows; donors as in D.
(D) Adherent RBCs against shear stress on laminin~$\alpha$5 (left) and HMEC-1 (right). Symbols, donor medians (CTR $n = 17$ and 12 donors, GD $n = 23$ and 18; bars, 68\% bootstrap intervals over donors). Red line, the CTR fit (reference runs, C); GD predictions with 1.3-fold Lu/BCAM, with the default bonds (dashed) and with $k_{\text{off}}^0/10$ (solid).
\end{figure}

\subsection*{Sickle-Cell Data Separate Receptor State from Receptor Number}
\begin{figure}[!tbp]
\begin{center}
\includegraphics[width=\textwidth,height=0.88\textheight,keepaspectratio]{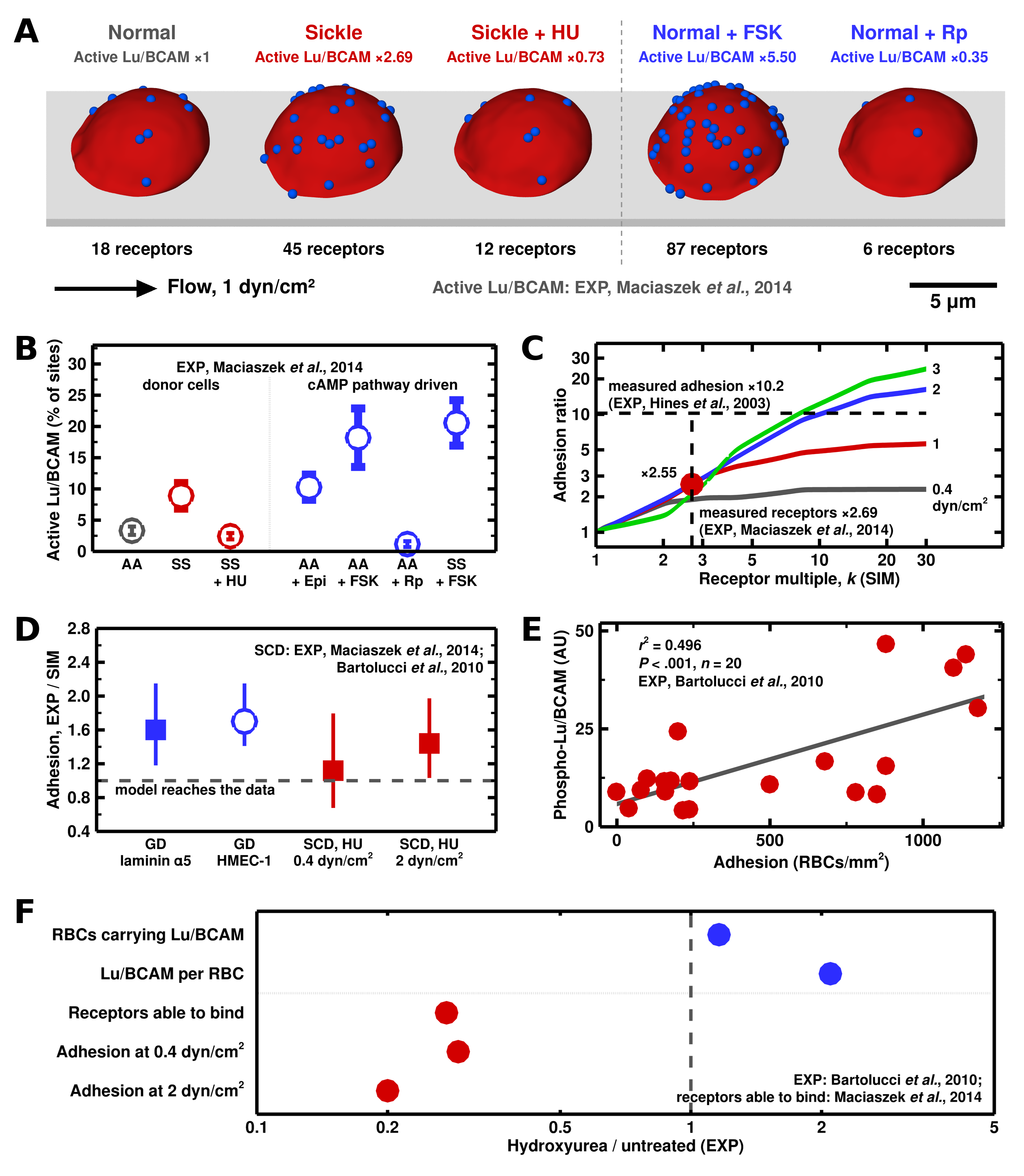}
\end{center}
\vspace{-0.15in}
\caption{\small{\bf Sickle-cell data separate receptor state from receptor number.}
Densities of untreated SS RBCs on laminin differ 13-fold between laboratories (200--2590~RBCs/mm\textsuperscript{2} at 1--2~dyn/cm\textsuperscript{2})~\cite{bartolucci2010decreased,hillery2000hydroxyurea}, so every quantity is a ratio within one study.}
\label{fig:scd}
\end{figure}

Lu/BCAM--laminin~$\alpha$5 adhesion is not peculiar to GD, and the receptor is measured more completely in sickle cell disease (SCD), where single-molecule force spectroscopy counts the receptors that can bind among those the cell carries~\cite{maciaszek2014sickle}. We applied the population model, fitted to our control donors and otherwise unchanged, to the published SCD measurements (Fig.~\ref{fig:scd}; Materials and Methods). The predictions needed no new simulation: our grid of shear stresses, 0.39--3.15~dyn/cm\textsuperscript{2}, brackets every shear stress used in those assays.

On normal RBCs, epinephrine, the agonist these cells meet in the circulation, raises the share of surface sites giving an unbinding event, a measure of binding-competent receptors, from 3.31\% to 10.27\%, after a 30-min incubation of mature RBCs~\cite{maciaszek2014sickle}. Driving the cAMP pathway further with forskolin raises it to 18.19\%, and blocking protein kinase~A with Rp-cAMPS lowers it to 1.17\%, each after the same incubation~\cite{maciaszek2014sickle}. Activation alone thus moves the number of active receptors over a 16-fold range (Fig.~\ref{fig:scd}B). Sickle cells are not already saturated: forskolin takes them from 8.89\% to 20.56\%. Untreated sickle (SS) RBCs carry 2.69-fold more active receptors than normal (AA) RBCs (8.89\% against 3.31\%; Fig.~\ref{fig:scd}A, B). Yet a single bond is as strong in SS as in AA cells: the unbinding force at a loading rate of 24,000~pN/s is 40.0 against 38.9~pN (fig.~S5). In SCD, measurement argues against a stronger bond, which Fig.~\ref{fig:alternatives} tests only by simulation.

Comparing sickle with normal cells is not a clean test of the model. SS RBCs adhere to laminin 10.2-fold more than AA RBCs at 1~dyn/cm\textsuperscript{2} (317 against 31~RBCs/mm\textsuperscript{2})~\cite{hines2003novel}, whereas the 2.69-fold rise in active receptors predicts only 2.55-fold. But at this shear stress no receptor number can reach the measurement: once every cell of the population is arrested the ratio saturates at 5.8-fold (Fig.~\ref{fig:scd}C), and the comparison crosses two laboratories.

Hydroxyurea gives the test with fewer receptors, far from that ceiling. It lowers the active-receptor count of SS RBCs 3.66-fold, from 8.89\% to 2.43\%~\cite{maciaszek2014sickle}, and over the same treatment adhesion to laminin falls 3.4-fold at 0.4~dyn/cm\textsuperscript{2} (550 to 160~RBCs/mm\textsuperscript{2}) and 5.0-fold at 2~dyn/cm\textsuperscript{2} (200 to 40)~\cite{bartolucci2010decreased}. The model predicts falls of 3.1- and 3.5-fold, so the measurement exceeds the prediction 1.12- and 1.44-fold (Fig.~\ref{fig:scd}D). The standard errors of the three active-receptor counts, each over 12 to 15 cells, give these ratios 68\% intervals of 0.68 to 1.80 and 1.03 to 1.97. The model reaches the measurement in 38\% and 14\% of the draws (Materials and Methods). Given the receptors that bind, the measured fall therefore does not exclude the model at the 5\% level. The counts are too uncertain to place these ratios below the 1.60 and 1.70 of GD. The prediction leaves out that the unbinding force of the single bond is also 12\% lower under hydroxyurea (35.1 against 40.0~pN; fig.~S5)~\cite{maciaszek2014sickle}. A weaker bond can only add to the predicted fall. A second laboratory reports a smaller fall on laminin under hydroxyurea, 1.53-fold, with no AA controls~\cite{hillery2000hydroxyurea}.

Bartolucci et al.~\cite{bartolucci2010decreased} separate receptor number from receptor state directly. Across 20 SS patients, adhesion to laminin rises with how much Lu/BCAM is phosphorylated ($r^2 = 0.496$, $P < .001$; Fig.~\ref{fig:scd}E). No correlation is found with how much Lu/BCAM the cell carries, measured either as the percentage of Lu/BCAM-positive RBCs or as their mean fluorescence intensity~\cite{bartolucci2010decreased}. The state can also change without the number on the timescale of an hour. Dibutyryl cAMP raises SS adhesion to laminin 2.5-fold within the hour with no change in the number of receptors the cell carries, and a protein kinase~A inhibitor blocks the rise~\cite{hines2003novel}. Soluble Lu/BCAM removes 96\% of the stimulated adhesion, so the Lu/BCAM bond is the one modulated~\cite{hines2003novel}. Hydroxyurea then moves number and state in opposite directions (Fig.~\ref{fig:scd}F): the receptors able to bind fall 3.66-fold in the cells of one study~\cite{maciaszek2014sickle}. In the patients of Bartolucci et al., adhesion falls while the total receptor count \emph{rises}, 1.16-fold by the percentage of Lu/BCAM-positive cells and 2.10-fold by their mean fluorescence intensity (both $P < .001$ over 19 untreated and 19 treated). In the five patients followed before and during treatment, what falls under hydroxyurea is the phosphorylation itself ($P = .005$)~\cite{bartolucci2010decreased}. Under hydroxyurea the cells thus carry more Lu/BCAM but fewer receptors able to bind, and they adhere less. Receptor number and receptor state are therefore not interchangeable, as the GD shortfall implies.

\begin{figure}[!t]
\small\noindent{\bf Figure~\ref{fig:scd} (continued).}
(A) The model cell as in Fig.~\ref{fig:shortfall}A, with the mean receptor fraction of the control fit on laminin~$\alpha$5 multiplied by the ratio of active receptors measured in five of the seven conditions of B. Receptor counts: 18, 45, 12, 87 and 6 (single draws); the five simulations differ only in $f$.
(B) Share of sampled surface sites giving a Lu/BCAM--laminin~$\alpha$5 unbinding event, a measure of the receptors able to bind, by atomic force microscopy (AFM)~\cite{maciaszek2014sickle}. AA, normal; SS, sickle; HU, hydroxyurea; Epi, epinephrine; FSK, forskolin; Rp, Rp-cAMPS. Left, cells as they come from the donor; right, the cAMP pathway driven or blocked by a 30-min incubation. Bars, SEM over 9--15 cells from 3--5 subjects.
(C) Adhesion ratio predicted by the model against the receptor multiple $k$, at four shear stresses, for the population fitted to our control donors on laminin~$\alpha$5 (Fig.~\ref{fig:calibration}). Dashed lines, the measured 2.69-fold rise in active receptors and the 10.2-fold rise in adhesion at 1~dyn/cm\textsuperscript{2}~\cite{hines2003novel,maciaszek2014sickle}; filled symbol, the model at the measured receptor rise.
(D) Measured over predicted adhesion in the two diseases. GD: on laminin~$\alpha$5 and on HMEC-1 (this work; bars, 68\% bootstrap intervals over donors and simulation runs). SCD: the hydroxyurea test at 0.4 and 2~dyn/cm\textsuperscript{2}, where receptors and adhesion were measured over the same treatment, in separate cohorts~\cite{maciaszek2014sickle,bartolucci2010decreased}. Bars on the SCD points, 68\% intervals from the published standard errors of the three receptor counts alone (Materials and Methods). Dashed line, a ratio of 1.
(E) Lu/BCAM phosphorylation against adhesion for 20 SS patients, read from Figure~4E of ref.~\cite{bartolucci2010decreased} (Materials and Methods); line, the regression over those points.
(F) The change of each quantity under hydroxyurea: its value in treated patients divided by its value in untreated patients. Dashed line, no change. A point to the right of it rises under hydroxyurea, and a point to the left falls. Blue, the Lu/BCAM the cells carry, which rises: the percentage of RBCs that carry Lu/BCAM and the number of Lu/BCAM molecules per RBC, estimated from the mean fluorescence intensity, 19 against 19 patients, both $P < .001$, read from Figures~2C and 2D of ref.~\cite{bartolucci2010decreased}. Red, the receptors able to bind and the adhesion, which fall: active sites by AFM, 12 against 15 cells~\cite{maciaszek2014sickle}, and adherent density at 0.4 and 2~dyn/cm\textsuperscript{2}, 34 against 28 patients~\cite{bartolucci2010decreased}.
\end{figure}

\subsection*{Stiffness Retains the Cell at the Splenic Slit while Adhesion Runs the Other Way}
For adhesion the GD membrane is at best neutral: the stiffer cell holds fewer bonds and loses them sooner (Fig.~\ref{fig:alternatives}A). At the interendothelial slit of the splenic sinus, about 1.5~$\mu$m across, the same stiffening acts the other way (Supplementary Materials, fig.~S6, and movies~S1 and~S2). A simulated cell with only the reduced surface area of the GD membrane spends 1.60 times as long in the slit as a control cell and still emerges in every run. By contrast, a cell with only the stiffer GD membrane never leaves the slit in six runs of 1215~ms. Some runs fail to pass at 14.2 and 18.9~$\mu$N/m, and from 28.4~$\mu$N/m upward no cell moves, so the threshold lies at three to six times the healthy modulus. The moderate-GD membrane is at nine times~\cite{chai2024silico}. In the model, the same cell can adhere less and yet be retained more: adhesion follows the receptors able to bind, retention the stiffness of the membrane, and neither stands in for the other.

\section*{Discussion}
GD RBCs adhere more than control RBCs to activated endothelium and to laminin~$\alpha$5, and the difference grows with the shear stress at which the cells are washed, significantly so on endothelium. We asked whether the higher Lu/BCAM expression of GD RBCs can account for this. A DPD model of single cells carrying stochastic Lu/BCAM--laminin bonds, averaged over a distribution of receptor numbers calibrated on the control donors, predicts the number of adherent cells as a function of receptor number. For the 1.3-fold rise in receptor number measured on mature GD RBCs, the prediction falls short of the GD data by a factor of 1.6 on laminin and 1.7 on HMEC-1. With 12 to 23 donors per group, the 95\% bootstrap interval of this factor excludes agreement on HMEC-1, whereas on laminin only the 68\% interval does. The result on HMEC-1 further assumes that GD and control cells reach the substrate in equal numbers (below). To match the data, the receptor number of every cell would have to rise 2.4-fold (laminin) or 4.6-fold (HMEC-1). These are lower bounds, since the data set no upper limit. The stiffer membrane of GD RBCs widens this gap, and a higher bond affinity narrows it mainly at low shear stress (Fig.~\ref{fig:alternatives}). We also applied the model unchanged to published measurements in sickle cell disease, where the receptors able to bind are counted directly and a single bond is as strong as in normal RBCs. Under hydroxyurea, adhesion falls while the receptors the cell carries rise. The model comes within 1.1- to 1.4-fold of this fall, a remainder that the uncertainty of the receptor counts can account for. There too, what the data ask for is a change of receptor state (Fig.~\ref{fig:scd}).

Both estimates lie above the 1.3-fold rise in receptor number, and the estimated 5.7-fold rise in phosphorylated receptors~\cite{franco2013abnormal} lies inside both of their 68\% intervals. Binding of Lu/BCAM to laminin~$\alpha$5 is controlled by phosphorylation of its cytoplasmic tail, and a receptor whose protein kinase~A site is blocked adheres far less~\cite{gauthier2005protein}. A GD cell may therefore carry more receptors and a larger share of receptors able to bind. A change in the state of the receptor could instead raise the affinity of each bond. In a reduced Bell description, a lower force sensitivity of the bond shifts a detachment curve almost as more receptors do (Supplementary Materials), but in the full model a higher affinity acts differently (Fig.~\ref{fig:alternatives}). Because binding is nearly saturated, a slower off-rate adds few bonds. It keeps lightly loaded bonds from dissociating at low shear stress, whereas at high shear stress most bonds still break at their rupture length. Its gain therefore fades as the shear stress rises, whereas the excess of GD adhesion grows. On laminin each of the three rate changes brings the model within the 68\% interval of the data. Yet the strongest, a tenfold lower off-rate, still falls 1.4- to 1.8-fold short at 1.0--1.6~dyn/cm\textsuperscript{2} on laminin and 1.7- to 2.2-fold short at 2 and 3~dyn/cm\textsuperscript{2} on HMEC-1. With each of the three rate changes, GD would still need 2.2 to 3.5 times the receptors of a control cell. Within the model, the remainder of $k^*$ thus behaves like more bonds, as expected if a GD cell carries a larger share of binding-competent receptors. Activation of the receptor remains the most direct candidate for it. The moderate correlation between Lu/BCAM level and adhesion to laminin across donors ($\rho = 0.64$)~\cite{franco2013abnormal} leaves room for such a second factor: the state of the receptor. Besides phosphorylation, this state includes the dimerization of Lu/BCAM and its link to spectrin. In epithelial (MDCK) cells Lu/BCAM promotes migration on laminin~521 only when it both forms dimers and is phosphorylated~\cite{guadall2019dimerization}. Lu/BCAM that can no longer bind spectrin stays longer at the surface of these cells and increases their adhesion and spreading on laminin~511/521~\cite{collec2011novel}. In sickle cell disease, phosphorylation of Lu/BCAM triggers the adhesion of reticulocytes to laminin, whereas dense RBCs adhere firmly without it~\cite{bartolucci2010decreased,lizarralde2021oxidative}. In these dense cells, oxidative stress changes the distribution of Lu/BCAM at the surface and its interaction with glycophorin~C, and this activates its adhesive function~\cite{lizarralde2021oxidative}.

If activation supplies the remainder, assays already applied to sickle cells should show it. Atomic force microscopy~\cite{maciaszek2014sickle} should count clearly more than 1.3 times as many active sites on GD as on control RBCs, 2.4 to 4.6 times by our central estimates. Forskolin raises the active sites of normal cells 5.5-fold and of sickle cells 2.3-fold. It should raise those of GD cells 1.6- to 3.0-fold. At the share reached on normal cells, GD cells with 1.3 times the receptors would hold 7.1 times the control count, above both estimates of $k^*$. Inhibiting protein kinase~A, which takes stimulated sickle-cell adhesion below its basal level~\cite{hines2003novel}, should lower GD adhesion at least 1.5-fold at 0.4~dyn/cm\textsuperscript{2} and, on HMEC-1, 2.3-fold at 3~dyn/cm\textsuperscript{2}, to within about 1.3-fold of control cells treated alike. A ratio of active sites near 1.3, or an excess that survives the inhibitor, would place the cause outside this pathway.

The analysis also shows what the adhesion curves cannot determine. They do not fix how widely the receptor number varies between cells, and they fix the mean receptor number only loosely. Lu/BCAM expression varies widely between GD samples~\cite{franco2013abnormal}, and in sickle cell disease the activation of Lu/BCAM differs between light and dense RBCs~\cite{lizarralde2021oxidative}. In the model, such differences between cells appear as a different spread of the binding-competent receptors, but a different spread in GD RBCs does not close the gap. At the 1.3-fold mean, the best width of the GD distribution on laminin is that of the controls. On HMEC-1, at the same mean, a narrower distribution matches the data at low shear stress but not at 2 and 3~dyn/cm\textsuperscript{2}. Where the receptors sit does not close it either: with the same 16 receptors gathered into patches of up to 2.2~$\mu$m, the mean arrest never rises at any shear stress tested, and only the spread between cells grows (Supplementary Materials, table~S4). Measuring Lu/BCAM and its phosphorylated form cell by cell would fix the width and test the mean directly.

Five simplifications bear on the size of the shortfall. First, the main analysis gives both groups the membrane of healthy RBCs, although GD RBCs are less deformable~\cite{franco2013abnormal} and have a higher shear modulus in our earlier models~\cite{chai2024silico,chai2026quantifying}. With the membrane of moderate GD the shortfall grows, from 1.67- to 2.24-fold on laminin and from 1.68- to 1.88-fold on HMEC-1 (Fig.~\ref{fig:alternatives}A, B; fig.~S3C). Second, the density of cells in contact with the substrate is taken from the controls. The groups do not differ significantly at the lowest shear stress on either substrate, which supports this choice without proving it. On HMEC-1 a GD density higher by the 1.43-fold ratio measured there would lower the shortfall to 1.19 and remove its significance (Supplementary Materials). The shortfall stays significant only for a density ratio below 1.21 there, and the donors bound that ratio to 0.97--1.85 (Supplementary Materials). Third, the bond kinetics are inherited from earlier models. The bond stiffness is set without fitting the GD data. The rule only asks the model to span the measured range of densities (Materials and Methods). A bond twice as stiff, run on the same spots, leaves the shortfall at 1.81-fold on laminin and 1.70-fold on HMEC-1 (fig.~S3C). The changes in the bond rates tested leave the shortfall in place at high shear stress (above).
 Fourth, the flow acts for at most 0.24~s in a run (Materials and Methods), whereas each washout of the experiment lasts three minutes. The arrest $A$ measures how firmly the bonds hold a cell under a given shear stress, and Eq.~\ref{eq:population} takes the density of cells that remain after a washout to be proportional to it. Detachment that develops over minutes is not resolved. The same assumption enters the calibration on the controls and the GD prediction. In runs ten times longer, 2.4~s, the arrest levels off after about 1~s, and 1.3 times the receptors raise it 1.06- and 1.27-fold at 0.79 and 2.36~dyn/cm\textsuperscript{2}, against 1.26- and 1.29-fold in the main runs (table~S5). Fifth, no data calibrate the predicted response to receptor number, and the one measurement of it is steeper than the model's (Results). The measured 1.3-fold rise accounts for the GD data only if adhesion responds to receptor number more steeply than either estimate. Narrowing the uncertainty of the shortfall needs more patients (Results).

A GD RBC can leave the circulation by two routes, and in the model they depend on different properties of it. Adhesion through Lu/BCAM depends on the receptors able to bind. Retention at a splenic slit depends on the shear modulus of the membrane and involves no receptor. With respect to stiffness the two run in opposite directions, so the same cell adheres less and is retained more, and a measurement of one gives no estimate of the other. The threshold we find for retention, between 14.2 and 28.4~$\mu$N/m, is three to six times the healthy shear modulus. It lies below the 42.6~$\mu$N/m of a moderate-GD membrane, which places GD cells clearly on the retained side of a passage that control cells make in 227~ms. It also distinguishes the two GD membrane changes: the reduced surface-to-volume ratio slows the passage 1.6-fold but never prevents it, whereas the stiffening alone prevents it. Whether a treatment restores passage should therefore be judged on the shear modulus. Cell shape does not decide it. The two routes also respond differently to treatment. Enzyme replacement therapy normalized the deformability and the morphology of GD RBCs but left their adhesion to laminin and their Lu/BCAM level unchanged, in 15 treated against 16 untreated patients~\cite{franco2017effect}. That study ascribes the unchanged adhesion in part to the short duration of the treatment.

Adhesion of RBCs to the vessel wall has been proposed to contribute to the ischemic events of GD~\cite{franco2013abnormal}. Our analysis suggests that the adhesion of GD RBCs depends on the activation state of Lu/BCAM as well as on its expression level. Phosphorylated Lu/BCAM is therefore a quantity to follow in patients and under treatment, alongside the deformability that governs the splenic route. The approach calibrates a single-cell bond model as a population on control samples and predicts patient samples from an independently measured molecular change. It applies to other disorders with abnormal RBC adhesion, such as polycythemia vera~\cite{wautier2007increased} and sickle cell disease. In the latter, the adhesion of single sickle cells under hypoxia has been followed in a microfluidic channel~\cite{papageorgiou2018simultaneous}, and adhesive DPD models already exist~\cite{LEI2012Quantifying,lei2013probing,Deng2019Quantifying}.

\section*{Materials and Methods}

\subsection*{Computational Approach}

We simulate single RBCs adhering to a ligand-coated wall under flow. We use DPD, one of the particle methods widely used for blood cells, soft matter and collective motion~\cite{fedosov2010multiscale,li2014erythrocyte,deng2020quantifying,toscano2026graft,chai2026pocked,chai2023dynamics,zhang2020deep,zhang2021deep}. The equations and every parameter value are in the Supplementary Materials. Snapshots were rendered with OVITO~\cite{stukowski2010ovito}.

\subsection*{Dissipative Particle Dynamics Model}

Blood plasma is treated as a mesoscopic fluid of coarse-grained particles, in the form of DPD given by Hoogerbrugge and Koelman~\cite{hoogerbrugge1992simulating} and by Groot and Warren~\cite{groot1997dissipative}. The three pairwise forces, which act in equal and opposite pairs so that momentum is conserved locally and the fluid is hydrodynamic~\cite{Fedosov2011Multiscale}, and the fluctuation--dissipation relation that fixes the temperature~\cite{groot1997dissipative}, are given in the Supplementary Materials. The fluid inside the cells is the same DPD fluid as outside. The DPD parameters are listed in table~S6.

The channel is periodic in the flow and vorticity directions and bounded by two flat walls. A uniform body force $f_x$ per particle on the fluid and on the cells drives the flow. Without cells, the shear stress at the walls is $\tau_w = \rho f_x H/2$, where $\rho = 4$ is the number density of the fluid and $H = 10$ the distance between the walls. Model quantities are converted to physical units as in our previous work~\cite{chai2026quantifying}: the unit of length is 1~$\mu$m and the unit of stress 0.394~dyn/cm\textsuperscript{2}, so that $\tau_w = 7.88\,f_x$~dyn/cm\textsuperscript{2}, and the unit of time, set by the solvent viscosity we measured in Poiseuille flow, is $t_0 = 2.2$ to 2.4~ms (Supplementary Materials).
\subsection*{RBC Model}

Adhesion is decided at the scale of one cell, so the cell has to deform as a real RBC does while the flow around it is still resolved. An RBC has no nucleus. It is a biconcave bag about 8~$\mu$m across whose membrane is viscous and elastic at once. That combination lets it be forced through capillaries as small as 3~$\mu$m~\cite{diezsilva2010shape,Fedosov2011Multiscale}. We represent the membrane by $N_v$ DPD particles at positions $X_i$ ($i \in 1,\ldots, N_v$), joined into a triangulated network whose edges each carry a spring. Its area and the volume it encloses are held fixed by constraint forces, which make the membrane and the cell incompressible, and a resistance to the angle between neighboring triangles gives the membrane its bending rigidity.

Both groups of cells are represented with the membrane of healthy RBCs: shear modulus $\mu_0 = 4.73~\mu\mathrm{N/m}$ and bending coefficient $k_b = 2.4 \times 10^{-19}~\mathrm{J}$~\cite{fedosov2010multiscale}, with the area, volume and other membrane parameters of table~S6. The model and its validation against optical-tweezers stretching and tank-treading in shear flow are described in our previous work~\cite{chai2024silico,chai2026quantifying}. We keep the healthy membrane for both groups in the main analysis, so that the comparison concerns the receptors alone, and test the shear modulus of moderate GD in our earlier model, 42.6~$\mu$N/m~\cite{chai2024silico}, as a variant (below).

\subsection*{Cell Adhesion Model}

Adhesion is represented by bonds that form and rupture stochastically between receptors on the cell membrane and ligands on the substrate~\cite{Fedosov2011Multiscale}. 

At each simulation time step $\Delta t$, a receptor within the capture distance $d_{\text{on}}$ of a free ligand binds it with probability $P_{\text{on}} = 1 - e^{-k_{\text{on}}\Delta t}$. Every existing bond ruptures with probability $P_{\text{off}} = 1 - e^{-k_{\text{off}}\Delta t}$ at each step, and unconditionally once it is stretched beyond the rupture length $d_{\text{off}}$. Both rates depend on the bond length $l$~\cite{dembo1988reaction}:

\begin{align}
k_{\text{on}} &= k_{\text{on}}^0 \exp\left(-\frac{\sigma_{\text{on}} (l - l_0)^2}{2k_\textrm{B}T}\right), \\
k_{\text{off}} &= k_{\text{off}}^0 \exp\left(\frac{\sigma_{\text{off}} (l - l_0)^2}{2k_\textrm{B}T}\right),
\end{align}
where $l_0$ is the equilibrium bond length, $k_\textrm{B}T$ is the thermal energy of the DPD fluid, and $\sigma_{\text{on}}$ and $\sigma_{\text{off}}$ set how strongly extension slows binding and accelerates rupture. A bond pulls its two particles together with the harmonic force $F_{\text{adh}}(l) = 2k_s (l - l_0)$, where $k_s$ is the bond spring constant, so that the bond stiffness $2k_s$ is 3.94~pN/$\mu$m at $k_s = 50$. We take $l_0 = 0$. Each receptor and each ligand carries at most one bond. A bond thus carries at most 2.96~pN. Single Lu/BCAM--laminin~$\alpha$5 bonds unbind at 39~pN in force spectroscopy at 24,000~pN/s (fig.~S5), and bond strength falls with the loading rate~\cite{evans1997dynamic}.

This adhesion model has been validated in simulations of RBC adhesion in malaria~\cite{Fedosov2011Multiscale,fedosov2011wall,Fedosov2011Quantifying} and sickle cell disease~\cite{LEI2012Quantifying,lei2013probing}. We keep its kinetic parameters $k_{\text{on}}^0$, $k_{\text{off}}^0$, $\sigma_{\text{on}}$, $\sigma_{\text{off}}$, $d_{\text{on}}$ and $d_{\text{off}}$ at the values of the implementation our code derives from (table~S6). Of these, $\sigma_{\text{on}}$ and $\sigma_{\text{off}}$ are rescaled so that $\sigma/k_\textrm{B}T$, and with it every rate, is unchanged at the temperature of our DPD fluid. The bond spring constant is selected by the criteria given below, and the quantity calibrated against experiment is the distribution of receptor number across cells.

\subsection*{Model Setup and Parameter Estimation}

We use the multiscale RBC (MS-RBC) model~\cite{fedosov2010multiscale, lei2013probing} with $N_v = 500$ vertices, surface area $A_0 = 132.87\,\mu\text{m}^2$ and enclosed volume $V_0 = 92.46\,\mu\text{m}^3$. The MS-RBC model has been validated in computational studies of blood flow under normal and pathological conditions~\cite{peng2013lipid,pivkin2016biomechanics,chai2026reticulocyte,chai2026stomatocyte}, including malaria~\cite{Fedosov2011Quantifying}, sickle cell anemia~\cite{lei2013probing, li2017patient}, hereditary spherocytosis~\cite{chang2016md}, and diabetes mellitus~\cite{CHANG2017Diabete,chang2018quantifying}.

\textbf{Adhesion simulations.} We apply this model to Lu/BCAM on the RBC membrane binding laminin~$\alpha$5 on the substrate~\cite{hammer1992simulation,fedosov2011wall}. Each simulation contains two RBCs with the healthy membrane parameters of table~S6, side by side above the lower surface of a channel that is periodic in the flow and vorticity directions, 40 by 20~$\mu$m (Fig.~\ref{fig:setup}B). This lower surface carries a layer of ligand particles, and the two cells are far enough apart that no ligand can reach both, so they count as two samples. Each of a cell's 500 membrane vertices carries a receptor with probability $f$. All vertices also feel a weak non-specific attraction to the substrate, which keeps the membrane within reach of the ligands (Supplementary Materials). A run has two phases. The first, $10^5$ time steps without flow, stands in for the perfusion of the experiment at low shear stress: the cells settle and their bonds reach a steady population. Then, for $10^5$ steps, a body force drives the flow at a constant shear stress $\tau_w$. In the loaded phase the dissipative force between membrane vertices and the particles of the substrate and of the wall beneath it is 100 times weaker than for all other pairs of particles. Thus only the bonds resist the shear (the exception at the lowest shear stress is given below).

Adhesion is read from how much the bonds slow a cell down. The steady velocity $v$ of each cell's center of mass along the flow is the least-squares slope of its position over the second half of the loaded phase. We compare it with the mean velocity $v_0$ of eighteen cells that cannot form bonds ($k_{\text{on}}^0 = 0$) at the same $\tau_w$. The arrest $A = 1 - v/v_0$ is 1 for a cell held in place and 0 for one that moves like a non-adherent cell. Intermediate values mean that the cell slides. We compute $A(\tau_w; f)$ at eight receptor fractions, $f = 1/64$, $1/32$, $1/16$, $3/32$, $1/8$, $1/4$, $1/2$ and $1$, and six shear stresses from 0.39 to 3.15~dyn/cm\textsuperscript{2} ($f_x = 0.05$ to 0.40). This range covers the experimental shear stresses used in the fits (below). Each point has six independent runs (twelve cells), and those at $f = 1/64$ twelve (twenty-four cells). For a given random seed, the receptors at a larger $f$ include those at every smaller $f$. In each run the layer of ligand particles is shifted by its own random offset in the plane of the wall, so that the two cells of every run settle over their own spots of the substrate. The models compared in Fig.~\ref{fig:alternatives} were run without this shift, so that in all of their runs the two cells settle over the same two spots. Each is compared with reference runs of the default model made the same way (Supplementary Materials).

\textbf{Bond spring constant.} We do not fit $k_s$. No cell has more receptors than at $f = 1$, so a population retains at most $c\,A(\tau_w; 1)$ adherent cells per unit area, where $c$ is the density of cells in contact with the substrate (below). The model can represent the GD data only if this ceiling lies well above them. We take the smallest $k_s$ on the grid 25, 50, 100, 200 that meets two conditions. (i) At the highest laminin shear stress a cell with $f = 1$ stays largely arrested: $A(1.6~\text{dyn/cm}^2; 1) \geq 2R$, where $R = N_{\text{GD}}(1.6)/N_{\text{CTR}}(0.3) = 0.335$ is the median GD density at 1.6~dyn/cm\textsuperscript{2} relative to the median control density at 0.3~dyn/cm\textsuperscript{2}. Since $c \geq N_{\text{CTR}}(0.3)$, $R$ bounds from above the mean arrest the GD population needs at 1.6~dyn/cm\textsuperscript{2}, and the condition keeps that requirement below half of the ceiling. (ii) The receptor fraction that just holds a cell at the lowest laminin shear stress lies within the simulated range. Taking the shear stress a cell withstands as proportional to its receptor number (the simulations later give about $f^{0.59}$), this fraction is $f^* \approx 0.3/\tau_{1/2}$, where $\tau_{1/2}$ is the shear stress (in dyn/cm\textsuperscript{2}) at which $A(\tau_w; 1) = 1/2$. We require $f^* \geq 1/32$. Both conditions concern only the response at $f = 1$ and the range of the data. Neither condition uses a fit to either group. They select $k_s = 50$, with $A(1.6~\text{dyn/cm}^2; 1) = 0.87$ and $f^* = 0.056$. The value $k_s = 25$ fails condition (i), with $A = 0.59$ (table~S7).

\textbf{Receptor distribution.} A model receptor is a binding site on a membrane vertex. It is not a single Lu/BCAM molecule, and only ratios of $f$ are compared with measured receptor levels. The number of receptors may differ from cell to cell, so we describe the RBCs of a group as a population whose receptor fractions follow a gamma distribution with mean $\bar f$ and coefficient of variation CV. Cells drawn above $f = 1$ are assigned $f = 1$. A small CV approaches a population of identical cells. We write the adherent density as
\begin{equation}
N(\tau_w) = c\,\mathbb{E}_f\!\left[A(\tau_w; f)\right].
\label{eq:population}
\end{equation}
A cell cannot be held less firmly at a lower shear stress or with more receptors. We therefore first replace the simulated table of $A$ by the closest table, in least squares weighted by the number of cells at each point, that is non-increasing in $\tau_w$ and non-decreasing in $f$ (an isotonic regression). This removes sampling noise and changes 6 of the 48 entries, by at most 0.083 (Fig.~\ref{fig:arrest_map}B).
Between simulated points, $A$ is interpolated linearly in $f$, with $A(\tau_w; 0) = 0$, and linearly in $\log\tau_w$. The expectation then has a closed form in terms of incomplete gamma functions. We fit $c$, $\bar f_0$ and $\mathrm{CV}_0$ to the median control curve of each substrate by least squares in $\log N$, with $c$ in closed form, over the shear stresses of at least 0.4~dyn/cm\textsuperscript{2}: nine on laminin and ten on HMEC-1. Lower shear stresses lie below the simulated range. The control medians held out there are compared with the fit in the Supplementary Materials.
 Because the control curves leave $\mathrm{CV}_0$ undetermined (Results), we repeat the analysis with $\mathrm{CV}_0$ fixed at 0.2, 0.5 and 1.0 and with a lognormal distribution (table~S8). Another variant computes $v_0$ without the eight non-adherent cells that crept at less than half the speed of the cell-free fluid at their height, all at the lowest shear stress (Supplementary Materials, fig.~S4).

The GD curves are analyzed with $c$ held at its control value. The density $c$ is set by the perfusion protocol, the same for both groups, and the two groups do not differ significantly at the lowest shear stress on either substrate (Results). We then compute (a) the curve predicted for the higher Lu/BCAM expression of GD RBCs, $\bar f = 1.3\,\bar f_0$ at unchanged CV. This prediction involves no fit to the GD adhesion data: 1.3 is the ratio of the median Lu/BCAM levels on mature GD and control RBCs in Figure~4A of Franco et al.~\cite{franco2013abnormal}. We also compute (b) the receptor multiple $k^*$ for which $\bar f = k^*\bar f_0$ fits the GD curve at unchanged CV, (c) the CV that fits it at $\bar f = 1.3\,\bar f_0$ and (d) a free fit of $\bar f$ and CV.

Adhesion simulations are run with a modified version of LAMMPS at a time step $\Delta t = 0.001$ in DPD units. A run of $2\times10^5$ steps (two cells) takes about 10 CPU core-hours. The loaded phase, $10^5$ steps or $100\,t_0$, lasts at most 0.24~s, far shorter than the three-minute washouts of the experiment.

\textbf{Model variants and bond-resolved runs.} To test the model's assumptions, we repeat the grid of receptor fractions and shear stresses with one change at a time: the membrane of moderate GD, 42.6~$\mu$N/m~\cite{chai2024silico}, or one third of the body force on it, a twice stiffer bond, and five changes of the bond kinetics. A GD membrane and altered bond kinetics are hypotheses about GD RBCs, so for those variants the control population is calibrated with the default model and only the GD prediction uses the changed one. To follow single bonds we rerun 15 conditions and estimate bond lifetimes under load with the Kaplan--Meier method. Each variant and each rerun is listed in the Supplementary Materials.

\subsection*{Validation of the Model Components}
Each ingredient of the model was tested before use: the solvent alone forms a parabolic velocity profile between the walls, the bond kernel reproduces the closed-form kinetics of a single receptor--ligand pair, cells that cannot bind are carried along by the flow, and a cell with a receptor on every membrane vertex stays largely arrested at the bond stiffness we use. The tests are described in the Supplementary Materials (fig.~S7).

\subsection*{Experimental Setup}

\textbf{Microfluidic Adhesion Assay:} Adhesion was quantified in a plate flow chamber on two substrates: microslides coated with purified laminin~$\alpha$5, a component of the sub-endothelial matrix, and confluent monolayers of TNF$\alpha$-activated human microvascular endothelial cells (HMEC-1). Fibronectin-coated microslides provided a specificity control. RBCs from GD patients and healthy controls were resuspended at 0.5\% hematocrit and perfused at 37$^\circ$C. On endothelium, cells were infused at 0.2~dyn/cm\textsuperscript{2}, flow was stopped for 45~minutes to permit adhesion, and adherent cells were then counted after successive three-minute washouts from 0.2 to 3~dyn/cm\textsuperscript{2}. On laminin, cells were perfused at 0.3~dyn/cm\textsuperscript{2} for five minutes, washed and counted at the same shear stress, and again after three-minute washouts from 0.4 to 1.6~dyn/cm\textsuperscript{2}. The observable is the absolute surface density of adherent RBCs (cells/mm\textsuperscript{2}) at each shear stress, averaged over the imaged fields of a given donor. We do not reduce each donor to a single detachment threshold, because a half-detachment shear stress fitted to these curves separates the groups by less than its own scatter.

\textbf{Ethics Statement:} We used data from experiments previously done in Franco et al.~\cite{franco2013abnormal}, approved by the local ethics committee in accordance with the principles of the Declaration of Helsinki (CPP [Personal Protection Committee] agreement, \^Ile-de-France~VI of the Hospital Group, obtained on 16 December 2010). No blood was collected for this work.

\subsection*{Published Data}

Two quantities used in Fig.~\ref{fig:scd} are plotted in ref.~\cite{bartolucci2010decreased} but not tabulated, and were recovered from the published figures, with checks in the Supplementary Materials. All other published values used there are summary statistics stated in the cited work.

The intervals on the SCD points of Fig.~\ref{fig:scd}D propagate the published standard errors of the three active-receptor counts~\cite{maciaszek2014sickle} through the model, as described in the Supplementary Materials.

The intervals on the 1.3-fold rise in Lu/BCAM and on the 5.7 come from resampling the donors read from Figures~4A and 4D of ref.~\cite{franco2013abnormal} (Supplementary Materials).

\subsection*{Statistical Analysis}
Group comparisons are two-sided Mann--Whitney tests. Fewer than 2.2\% of the values of one donor at one shear stress reach zero adherent cells on either substrate, so these comparisons are not set by a measurement floor. We summarize the agreement of each model with the GD data by the ratio of the GD donor medians to the model, averaged geometrically over the fitted shear stresses, so that a ratio of 1 means agreement on average. Confidence intervals come from bootstrap replicates~\cite{efron1979bootstrap} that resample donors within each group and the simulations at each grid point. The table of each replicate is made monotone as above. For the shortfall and $k^*$ of the main fits, 10,000 replicates draw whole runs and refit the controls from five starting points. For the control fits, the variants of fig.~S4 and the models run on two fixed spots, 500 replicates draw single cells (5000 for fig.~S3C and table~S3). Replicates with a resampled donor median of zero or without a control fit are skipped (at most 0.4\%). The share of replicates with a ratio of at most 1 measures how often the model reaches the data. Other resampling units give nearly the same intervals (table~S1). The shortfall on laminin is 1.64 with the five questioned runs included (table~S9). The significance of the shortfall does not come from a bootstrap tail. It is assessed by a permutation test over donors. If the model is right, the GD donor curves divided by the predicted GD/control ratio have the distribution of the control curves, so the donor curves of the two groups are exchangeable. We divide them so and recompute the shortfall for 200,000 random relabellings, with the calibrated model held fixed (table~S1). Whole curves are relabelled, so the correlation between shear stresses is kept. Dividing the GD curves by a further factor and repeating gives the largest ratio of the GD to the control density of cells in contact at which the shortfall stays significant.

\section*{Supplementary Materials}
This PDF file includes: Supplementary Text, Figs. S1 to S7, Tables S1 to S13, Legends for movies S1 and S2, References~\cite{espanol1995statistical,lei2010direct,macdonald1987kinetics,chai2022periodic}.

Other Supplementary Material for this manuscript includes the following: Movies S1 and S2.

\section*{Acknowledgments}
Simulations were carried out at the Center for Computation and Visualization of Brown University.
\textbf{Funding:} Research reported in this publication was supported by the National Heart, Lung, and Blood Institute (NHLBI) of the National Institutes of Health under award numbers R01HL154150 and 3R01HL154150-06. The award, including supplement 3R01HL154150-06, totals \textdollar{}1,269,793 to date, with 100\% from federal sources. The content is solely the responsibility of the authors and does not necessarily represent the official views of the National Institutes of Health. We also acknowledge support from the France 2030 program through the Idex Universit\'e Paris Cit\'e (ANR-18-IDEX-0001, GR-Ex).
\textbf{Author contributions:} Conceptualization: ZC, GEK. Funding acquisition: PAB, GEK. Investigation: ZC (simulations), MF (experiments). Supervision: GEK.
\textbf{Competing interests:} The authors declare that they have no competing interests.
\textbf{Data and materials availability:} All data needed to evaluate the conclusions are in the paper and the Supplementary Materials: the adhesion of every donor measured in the source experiments of Franco et al.~\cite{franco2013abnormal}, the runs left out of it, the readings taken from the published figures (tables S10 to S13), and every parameter of the model and the analysis. The code is an in-house extension of LAMMPS, available from the corresponding author for academic use.

\bibliography{ref_main}

\end{document}